\documentclass[aps,showpacs,twocolumn,superscriptaddress,prb,floatfix, 10pt, longbibliography]{revtex4-2}

\usepackage[utf8]{inputenc}
\usepackage{amsfonts}
\usepackage{amsmath}
\usepackage{amssymb}
\usepackage{bbm}
\usepackage{amsthm}
\usepackage{empheq}
\usepackage{enumitem}
\usepackage{verbatim}
\usepackage{color}
\usepackage{subfigure}
\usepackage[breaklinks]{hyperref}
\usepackage{graphicx}
\usepackage{times}

\begin{document}

\title{Theory of non-resonant Raman scattering from electrons in nodal and flat bands}
\author{Predrag Nikoli\'c}
\affiliation{Department of Physics and Astronomy, George Mason University, Fairfax, VA 22030, USA}
\affiliation{Institute for Quantum Matter at Johns Hopkins University, Baltimore, MD 21218, USA}
\date{\today}

\begin{abstract}

Raman scattering is emerging as a surprising probe of electron topology in quantum materials. It has been used recently to detect and characterize a topological phase transition that accompanies the magnetic transition in Nd$_2$Ir$_2$O$_7$. Here we present a theory of Raman scattering from nodal electrons with Weyl and quadratic band touching spectra, which has to reach beyond the standard effective mass approximation. After reviewing and providing the details of our previous theory development, we discuss several new results. We show that the light-polarization dependence of Raman scattering is universal in the case of Weyl electrons and given by an analytic expression, while it contains symmetry-protected features in the case of quadratic band-touching nodes. We also analyze modifications of the Raman signal due to the ubiquitous tilting of the Weyl spectrum, and argue that universality is lost only in a finite frequency range that springs out of the threshold frequency for untilted nodes. Finally, we explore the frequency dependence of Raman scattering for the case of Dirac electrons coexisting with a flat band in the same region of the first Brillouin zone, which is inspired by the material V$_{1/3}$NbS$_2$.

\end{abstract}

\maketitle

\section{Introduction}\label{intro}

Topology has become one of the most important properties of quantum materials in recent years. It is both a fundamental paradigm for classifying the states of matter \cite{Schneider2008, Kitaev2009, Gu2010, Chen2011, Bi2013, Schnyder2014, Senthil2014, Ryu2016, Wen2019a, Wen2019}, and a physical mechanism that protects the response of materials to external perturbations against fluctuations and disorder, which can be harnessed in technological applications \cite{Fu2008, Alicea2011, Alicea2012, Yan2022, Liu2023, Movafagh2025}. Topology refers to the features of the system, mathematically represented by topological invariants, which remain unaltered under arbitrary gradual changes of the system \cite{Nakahara1990, StoneGoldbart2009}. By this definition, no local probe can detect the topology of a material in experiments. This is a fundamental problem for the experimental characterization of topological materials. Quantum Hall systems are famously known for their topologically-protected and quantized transverse conductivity \cite{Klitzing1980, Tsui1981, Laughlin1981, Halperin1982, Thouless1982}, but other topological materials do not exhibit as pristine transport properties \cite{Konig2007, Roushan2009}, and spin liquids, in particular, are just as famously difficult to identify without ambiguity \cite{Savary2016}. Nodal semimetals are another vast arena of topological materials \cite{Ari2010, Pesin2010, Wan2011, Ran2011, Krempa2012, Burkov2012, Savary2014, Armitage2018} where experimental discovery \cite{Valla2014, Ong2015a, Ong2016, Borisenko2015, Kuroda2017, Shekhar2018, Zhang2018, Lei2018, Schoop2018, Suzuki2019} and theoretical predictions \cite{Vishwanath2014, Jho2013, Zhang2014, Nandkishore2014, Nomura2014, Li2015, Bardarson2016, Morimoto2016, Lai2017, Burkov2020a, Narang2020a, Nikolic2020a, Nikolic2020b, Burkov2023} are both pushing the frontier. Nevertheless, detailed identification of topology and symmetry protected Weyl nodes or quadratic band touching is still usually entrusted to band-structure calculations \cite{Onoda2015, Yang2017, Felser2017b, Felser2018b}. Experimental techniques for identifying topology are obviously crucial when theoretical methods have limited accuracy, e.g. in materials where rare earth atoms or strong correlations are present. Such materials are increasingly getting attention, since the magnetism of several magnetic Weyl semimetals can be plausibly related to their electron topology \cite{Gaudet2021, Tafti2021b, Tafti2023, Tjernberg2023, Boothroyd2023, Lygouras2025}.

Here we discuss the ability of photon Raman scattering to probe the features of an electron spectrum that emerge as a result of the topological and symmetry protection mechanisms \cite{Ari2010, Moon2013}. Raman scattering is as local or non-local probe as angle-resolved photoemission spectroscopy (ARPES). It does its work in momentum space, so in principle it can couple to the electrons at the putative nodal points in the spectrum and hence discern aspects of their coherent dynamics across large length scales in clean systems. Apart from obvious contamination of the Raman signal by non-nodal, non-electronic and incoherent sources, the main limitation of Raman scattering is posed by negligible momentum transfers between photons and electrons. Specifically, unlike ARPES, Raman scattering cannot detect spin-momentum locking, which is a defining property of the  Weyl spectrum. Nevertheless, it is somewhat surprising that this handicap also empowers Raman scattering with an exceptional sensitivity. The frequency dependence of the Raman scattering cross-section reflects the nodal electrons' density of states across an extended energy range, which in turn is controlled by the presence of nodes. When the nodes are created or annihilated in phase transitions, driven by the onset of time-reversal symmetry breaking in magnets, changes of the spectrum across small momentum scales in the first Brillouin zone are seen by photons as dramatic changes across, for them, large momentum scales. For this reason, the Raman-detected topological phase transition between quadratic band touching and Weyl spectra in Nd$_2$Ir$_2$O$_7$ looks discontinuous even though it is driven by a continuous magnetization transition at a critical temperature \cite{Nikolic2022a}.

Raman scattering has been traditionally used as a probe of phonons and collective excitations in materials \cite{Cardona2005, Cottam1986, Yang2021, Muhammad2025}, including Weyl semimetals where it provided the evidence of satisfied symmetry requirements for Weyl nodes \cite{Ding2016, Yan2016, Chen2018, Burch2021, Smith2020, He2024}. Its use as probe of electron dynamics is not as widespread, but not new either \cite{Deveraux2007, Wang2008b, Kashuba2009, Gallais2016}. Generally, itinerant electrons give rise to an incoherent background in the frequency dependence of the Raman scattering cross-section, which is usually undesirable when coherent phonon or magnon peaks are of interest. However, the precise frequency and polarization dependence of this background, which can be fitted to models, contains a great amount of useful information. In typical metals, the electronic component of the Raman signal is dominated by thermal and quantum fluctuations. The latter includes interaction effects that give electronic excitations a finite lifetime even at zero temperature. This has motivated both theoretical \cite{Deveraux2007} and experimental explorations of correlated systems such as cuprate high-$T_c$ superconductors with Raman scattering. An extremely useful theoretical tool for this purpose is the effective mass approximation since it summarizes into a simple formula the photon-induced virtual transitions between all energy levels in the electron spectrum. In recent years, Dirac materials such as graphene \cite{Wang2008b, Kashuba2009, Gallais2016}, and then Weyl semimetals \cite{Nikolic2022a}, have also been scrutinized as systems amenable to Raman studies. Here, the new feature is that interactions and fluctuations are not essential for the emergence of the Raman signal, but then the effective mass approximation cannot capture the dominant contribution.

Here we present a theory of Raman scattering for electrons with Weyl and quadratic band touching spectra. A review of our earlier work \cite{Nikolic2022a} is weaved in throughout the discussion in order to introduce detailed derivations of key results. Some aspects of this theory were developed for the purpose of interpreting the Raman experiments on Nd$_2$Ir$_2$O$_7$ and Pr$_2$Ir$_2$O$_7$. In the Nd compound, Raman scattering has revealed a topological phase transition from a Luttinger (quadratic band-touching) to a Weyl semimetal state at the critical temperature for the onset of magnetic order. The experiment is consistent only with a Luttinger semimetal at low temperatures in the Pr compound, but there a magnetic order is absent as well \cite{Machida2010}. This picture, afforded by the theoretical interpretation, is in line with other experiments and theoretical approaches \cite{Machida2010, Ari2010, Kondo2015, Nakayama2016, Goswami2016a, Armitage2017b, Roy2021, Ohtsuki2019, Xu2023}. Its possible relevance to pyrochlore iridates and topological quantum materials in general motivates the present work. 

Going beyond the review, we obtain several new results in order to handle important properties of realistic materials. First, we discuss in detail the dependence of the nodal-electron Raman signal on the photon polarization. We derive an analytical expression for the universal polarization dependence in the case of Weyl electrons, and discuss symmetry-forged features of the polarization dependence in the case of quadratic band touching. This information can be experimentally extracted to complement the characterization of the nodal electron spectrum from the frequency dependence of the Raman signal -- revealing the properties such as the Fermi energy relative to the node energy, the strength of the spin-orbit coupling, and even the strength of interactions or amount of disorder. Furthermore, we analyze the corrections to the Raman scattering from Weyl electrons caused by the tilting of type-I Weyl spectra. Only the special case of no tilt was analyzed before \cite{Nikolic2022a}, even though tilting is an ubiquitous feature in materials. We show that the main effect of tilting is the appearance of a finite range of frequencies within which Raman scattering looses universal features. The relevant threshold frequencies can be directly experimentally visualized only in very clean samples with long electron lifetime, otherwise they may be obtained by fitting the frequency dependence of the Raman signal to the formulas we derive. The knowledge of these threshold frequencies then provides an estimate of the tilt parameter in the model Hamiltonian of Weyl electrons. Finally, we consider additional features beyond the effective mass approximation in the Raman scattering forged by a Dirac band and a flat band. We show that a Dirac spectrum coexisting with a flat band in the same region of the first Brillouin zone gives rise to a specific recognizable feature in the Raman signal. This is inspired by models of metals on the kagome lattice, as well as the recently studied material V$_{1/3}$NbS$_2$ \cite{Ray2025, Ghosh2025}.

The paper is organized as follows. We begin by laying out the foundation of electronic Raman scattering in Sec.\ref{elRaman}. Then, we present a detailed derivation of the Raman cross-section in the case of Weyl electrons in Sec.\ref{Weyl}, and quadratic band touching in Sec.\ref{QBT}. The combination of a Dirac and a flat band is explored in Sec.\ref{Flat}. All conclusions and final discussions are summarized in Sec.\ref{Concl}. We use the units $\hbar=1$ and, at times, Einstein's notation for summations over repeated indices. 

\raggedbottom

\section{Electronic Raman scattering}\label{elRaman}

\begin{figure}
\subfigure[{}]{\includegraphics[height=0.9in]{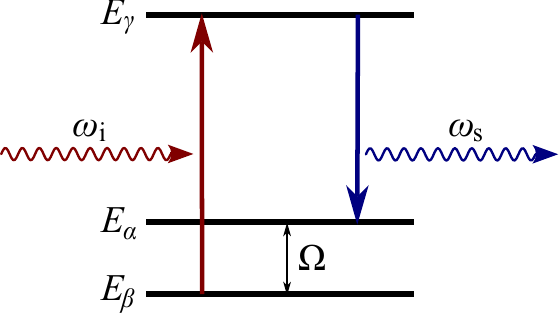}}\hspace{0.2in}
\subfigure[{}]{\includegraphics[height=0.5in]{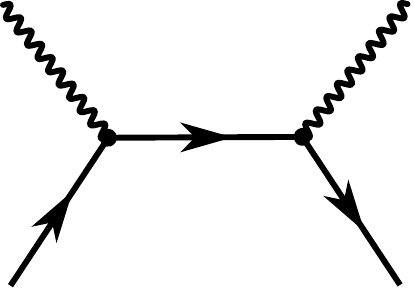}}\hspace{0.2in}
\subfigure[{}]{\includegraphics[height=0.5in]{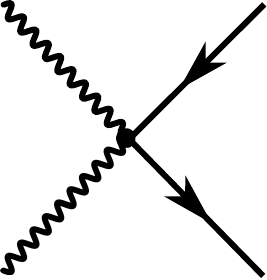}}
\caption{\label{RamanProcess}(a) Electron transitions induced by Raman scattering. $E_\alpha$, $E_\beta$ are the final and initial electron states respectively, $E_\gamma$ is an intermediate state; $\omega_{\textrm{i}}$, $\omega_{\textrm{s}}$ are the incoming and scattered photons' frequencies respectively, and their difference $\Omega=\omega_{\textrm{i}}-\omega_{\textrm{s}}$ is known as Raman shift frequency. (b,c) Feynman diagrams for photon-electron scattering that impacts the Raman cross-section. Wavy lines represent photons and solid lines with arrows represent electrons. (b) A process that involves a virtual intermediate state in general non-resonant Raman scattering (present in both relativistic and non-relativistic theories). (c) A ``diamagnetic'' process is present only in non-relativistic theories and not relevant for Raman scattering on Weyl/Dirac electrons with infinite lifetime.}
\end{figure}

Raman scattering is typically understood as photon scattering which temporarily excites an intermediate state of high energy. In resonant Raman scattering, the intermediate state is a real excitation produced by absorbing the incoming photon; this excitation quickly relaxes to a low-energy final state by emitting an outgoing photon, see Fig.\ref{RamanProcess}(a). Quantum mechanics also allows virtual intermediate states whose energy and momentum are off the mass-shell, and hence forbidden as enduring excitations; this supports non-resonant Raman scattering. Fig.\ref{RamanProcess}(b,c) illustrates the basic electronic photon scattering processes which contribute to the measured Raman scattering cross-section in experiments.

The foundation for the linear response theory of electronic Raman scattering is laid out in Ref.\cite{Deveraux2007}. We will adapt it here first to zero temperature and absence of interactions, since our main goal is to understand the Raman scattering on nodal electrons in semimetals, driven by the generation of particle-hole excitations. Later, all interaction and even thermal effects will be modeled with a simple self-energy correction in the zero-temperature formalism, which imparts a finite lifetime to electronic excitations. This is a dramatic approximation, but it qualitatively captures the most important physical effects while allowing analytical development of the theory.

The electronic contribution to the Raman differential scattering cross-section can be computed from \cite{Deveraux2007}:
\begin{equation}\label{CrossSection1}
\frac{\partial^{2}\sigma}{\partial\Omega\partial\omega_{\textrm{s}}} =
  \frac{\omega_{\textrm{s}}}{\omega_{\textrm{i}}} \left(\frac{e^{2}}{mc^{2}}\right)^{2} R(\Omega) \ , % m needs to be an effective mass, cancels out when m->inf
\end{equation}
where $e$, $m$ are electron's charge and mass, $c$ is the speed of light, $\omega_{\textrm{i}}$ and $\omega_{\textrm{s}}$ are the energies (frequencies) of the incident and scattered photons respectively, and $\Omega = \omega_{\textrm{i}}-\omega_{\textrm{s}}$ is the Raman shift frequency, i.e. the energy transferred from a photon to the electrons. The quantity
\begin{equation}\label{Rate1}
R(\Omega) =  -\frac{1}{\pi} \lim_{{\bf q}\to 0} \textrm{Im} \left\lbrace \chi({\bf q},\Omega)\right\rbrace
\end{equation}
is the scattering ``rate'' expressed in terms of the response function
\begin{equation}\label{Chi1}
\chi({\bf q},\Omega) = -i\sum_{{\bf k}\omega}\textrm{tr}\Bigl\lbrace G({\bf k},\omega)\gamma_{{\bf k}+{\bf q}}G({\bf k}+{\bf q},\omega+\Omega)\gamma_{\bf k}\Bigr\rbrace \ .
\end{equation}
When we later formulate the sum over wavevectors ${\bf k}$ as an integral, all quantities introduced in Eq.\ref{CrossSection1}--\ref{Chi1} will turn into corresponding densities. For our purposes, the Green's functions $G({\bf k},\omega)$ will describe quasiparticles with either a Weyl spectrum or quadratic band touching, having generally a finite lifetime $\tau=\Gamma^{-1}$ due to interactions, disorder, etc. The quasiparticle states $|\alpha\rangle$ with such nodal spectra carry a spin or band index in addition to the conserved momentum, so the Green's functions are generally matrices. The bare Green's function matrix for infinite lifetime can be conveniently written in terms of the non-interacting Hamiltonian matrix in momentum space:
\begin{equation}\label{Green1}
G_0({\bf k},\omega)=\frac{1}{\omega-H_{{\bf k}}+i\,\textrm{sign}(\omega)} \ .
\end{equation}
The Raman vertex function is similarly represented with a matrix $\gamma_{\bf k}$ whose matrix elements \cite{Deveraux2007}
\begin{eqnarray}\label{Gamma1}
&& \!\! \gamma_{\alpha\beta}^{\textrm{i,s}} = \langle\alpha| \gamma_{\bf k}^{\phantom{x}} |\beta\rangle = \hat{{\bf e}}_{\textrm{i}}^{\phantom{x}}\hat{{\bf e}}_{s}^{\phantom{x}}\rho_{\alpha\beta}({\bf q}_{\textrm{i}}-{\bf q}_{\textrm{s}}) +\frac{1}{m}\sum_{\gamma} \\
&& \times\!\left\lbrack \frac{p_{\alpha\gamma}^{\phantom{x}}(\hat{{\bf e}}_{\textrm{s}}^{\phantom{\dagger}},-{\bf q}_{\textrm{s}}^{\phantom{\dagger}})p_{\gamma\beta}^{\phantom{x}}(\hat{{\bf e}}_{\textrm{i}}^{\phantom{\dagger}},{\bf q}_{\textrm{i}}^{\phantom{\dagger}})}{E_{\beta}-E_{\gamma}+\omega_{\textrm{i}}}+\frac{p_{\alpha\gamma}^{\phantom{x}}(\hat{{\bf e}}_{\textrm{i}}^{\phantom{\dagger}},{\bf q}_{\textrm{i}}^{\phantom{\dagger}})p_{\gamma\beta}^{\phantom{x}}(\hat{{\bf e}}_{\textrm{s}}^{\phantom{\dagger}},-{\bf q}_{\textrm{s}}^{\phantom{\dagger}})}{E_{\beta}-E_{\gamma}-\omega_{\textrm{s}}}\right\rbrack \nonumber
\end{eqnarray}
are defined with respect to the quasiparticle's initial state $|\beta\rangle$ at momentum ${\bf k}$ and the final state $|\alpha\rangle$ at momentum ${\bf k}+{\bf q}$. This describes a virtual two-photon process: an incident photon with momentum ${\bf q}_{\textrm{i}}$ is absorbed by an electron and then a scattered photon is emitted at momentum ${\bf q}_{\textrm{s}}$, transferring ${\bf q}={\bf q}_{\textrm{i}}-{\bf q}_{\textrm{s}}$ to the electron gas. The first term of the Raman vertex function $\gamma_{\alpha\beta}^{\textrm{i,s}}$ is ``diamagnetic'', and the second term involves a summation over intermediate states $|\gamma\rangle$ which can belong to the high energy spectrum. The unit-vectors $\hat{{\bf e}}_{\textrm{i}}$ and $\hat{{\bf e}}_{\textrm{s}}$ specify the linear polarizations (electric field directions) of the incident and scattered photons respectively. $E_\alpha$ is the quasiparticle's energy in the state $|\alpha\rangle$, and
\begin{eqnarray}
\rho_{\alpha\beta}^{\phantom{x}}({\bf q}) = \rho_{\beta\alpha}^{*}(-{\bf q}) &=& \langle\alpha|e^{i{\bf q}{\bf r}}|\beta\rangle \\
p_{\alpha\beta}^{\phantom{x}}(\hat{{\bf e}},{\bf q}) = p_{\beta\alpha}^{*}(\hat{{\bf e}},-{\bf q}) &=& \hat{{\bf e}}\,\langle\alpha|e^{i{\bf q}{\bf r}}(-i\boldsymbol{\nabla}-\sigma^{a}{\bf A}^{a})|\beta\rangle \nonumber \ .
\end{eqnarray}
Since the vertex function is derived from the second order perturbation theory with respect to the electron-photon coupling, the matrix elements of the momentum operator $-i\boldsymbol\nabla$ are included in $p_{\alpha\beta}$. A new ingredient in this theory is the SU(2) gauge field ${\bf A}^a$ contracted into the Pauli matrices $\sigma^a$, $a\in\lbrace x,y,z \rbrace$. This gauge field is a mathematical encoding of the spin-orbit coupling that gives rise to the Weyl spectrum. The U(1) gauge invariance requires that the full operator $-i\boldsymbol{\nabla}-\sigma^a{\bf A}^a$ be coupled to the photons' U(1) gauge field ${\bf a}$.

The calculation of the vertex function can be complicated, so a popular approximation is to neglect the photon momenta ${\bf q}_{\textrm{i}}, {\bf q}_{\textrm{s}}$ and the momentum transfer ${\bf q} = {\bf q}_{\textrm{i}}-{\bf q}_{\textrm{s}}$ next to the electron momenta ${\bf k}$ in initial and final states $|\alpha,\beta\rangle$. This is justified due to $c\gg v$, where $v$ is the electrons' Fermi velocity. Furthermore, when the intermediate states $|\gamma\rangle$ in (\ref{Gamma1}) are sampled from the high-energy spectrum, their energies $E_\gamma$ are routinely much larger than the photon energies $\omega_{\textrm{i,s}}$, so that one can neglect $\omega_{\textrm{i,s}}$ in the vertex function. These measures lead to the ``effective mass approximation'' for the Raman vertex:
\begin{equation}\label{Gamma1b}
\gamma_{{\bf k}}^{\phantom{x}} = m\,\hat{e}_{\textrm{i}}^{a}\hat{e}_{\textrm{s}}^{b}\,\frac{\partial^{2}H_{{\bf k}}}{\partial k^{a}\partial k^{b}} \ .
\end{equation}
While generally very useful, the effective mass approximation trivially leads to $\gamma_{\bf k}=0$ for Weyl electrons because their low-energy Hamiltonian is linear in momentum, $H_{\bf k} \sim v \boldsymbol{\sigma}{\bf k}$. The effective mass approximation can pick contributions from the high-energy parts of the Weyl spectrum when the dispersion begins to deviate from the linear form $\epsilon_{\bf k}=v|{\bf k}|$; this is analyzed in Appendix \ref{appWeyl}. However, the essence of the photon scattering by Weyl electrons cannot be captured with the effective mass approximation, and we will need to carry out more accurate calculations.

\section{Weyl electrons}\label{Weyl}

The simplest Weyl spectrum obtains from the non-interacting Hamiltonian
\begin{equation}\label{WeylHamiltonian}
H_{{\bf k}} = \frac{k^{2}}{2m}+v\sigma^{a}k^{a}-\mu = \frac{({\bf k}-\sigma^{a}{\bf A}^{a})^{2}}{2m}-\mu' \ ,
\end{equation}
where an SU(2) gauge field
\begin{equation}\label{GaugeField}
{\bf A}^{a} = -mv\,\hat{\bf x}_{a}
\end{equation}
is introduced to capture the spin orbit coupling and produce a Weyl spectrum. The original chemical potential
\begin{equation}\label{ChemPot}
\mu = \mu'-\frac{3}{2}mv^{2}
\end{equation}
is the Fermi energy relative to the node energy. The Hamiltonian eigenstates $|\alpha\rangle\equiv|{\bf k},\sigma\rangle$, with $\sigma=\pm1$ are the eigenstates of momentum and the momentum-aligned spin projection. The corresponding eigenvalues are
\begin{equation}\label{WeylEnergy1}
E_{{\bf k}\sigma} = \frac{k^{2}}{2m} + \sigma vk - \mu \ .
\end{equation}
We include the quadratic correction in the Weyl electron dispersion, involving an effective mass $m$, in order to take advantage of the existing theory of Raman scattering developed for non-relativistic electrons \cite{Deveraux2007}. This also adds more realistic features to the spectrum. The price to pay is the formal existence of two Fermi surfaces for generic $\mu$, a ``small'' and a ``large'' one. The ``small'' Fermi surface belongs to the Weyl spectrum of interest, and the ``large'' one has the Fermi wavevector $k_{\textrm{F}}>mv$ which is comparable to or larger than the cut-off momentum for a realistic material. We will effectively ignore the dynamics at such large wavevectors since the present model does not capture it realistically.

\begin{figure}
\subfigure[{}]{\includegraphics[height=1.3in]{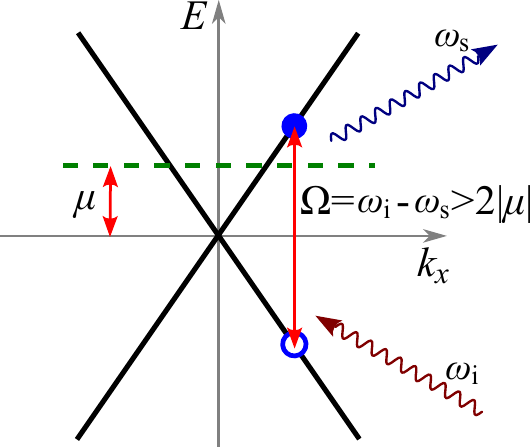}}\hspace{0.1in}
\subfigure[{}]{\includegraphics[height=1.3in]{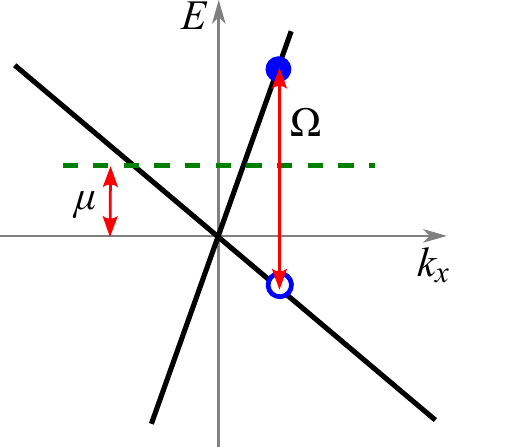}}
\caption{\label{RamanWeyl}(a) Illustration of the non-resonant Raman scattering from Weyl electrons with infinite lifetime and a spherically-symmetric spectrum. Incoming (i) and scattered (s) photons, represented by wavy lines, transfer energy $\Omega$ and negligible momentum to an electron-hole excitation. If the Fermi level is at energy $\mu$ relative to the node, than the minimum possible energy transfer with zero momentum transfer is the threshold frequency $\Omega=2|\mu|$ for the Raman scattering. The physical process is depicted in Fig.\ref{RamanProcess}, but the intermediate state is virtual and lives within the low-energy Weyl spectrum, hence cannot be handled by the effective mass approximation. (b) An illustration of Raman scattering from a ``tilted'' type-I Weyl node.}
\end{figure}

In the context of a realistic Weyl semimetal, the Hamiltonian (\ref{WeylHamiltonian}) serves as an effective theory for the low-energy Weyl quasiparticles located near a particular node wavevector ${\bf Q}$ in the first Brillouin zone. Therefore, ${\bf k}$ is the crystal momentum relative to ${\bf Q}$, and the parameters $v, m, \mu$ have values specific to the given node. This effective Hamiltonian can be made more realistic by introducing a fixed vector ${\bf a}$ into the kinetic energy term, as $({\bf k}-{\bf a}-\sigma^a{\bf A}^a)^2/2m$. Being formally a U(1) gauge field without flux, ${\bf a}$ adds a ${\bf Q}$-dependent ``tilt'' to the ``cone'' of the Weyl spectrum. For the sake of simplicity, we will initially idealize the Weyl spectrum and make it perfectly spherically symmetric by setting ${\bf a}=0$. Afterwards, we will identify the physical consequences of having ${\bf a}\neq 0$. Furthermore, it is important to note that gauge invariance requires the U(1) gauge field of the photons to couple minimally to the momentum ${\bf k}$ in the effective Hamiltonian (\ref{WeylHamiltonian}) -- in other words, it must be a fluctuating part of ${\bf a}$. The fundamental formulas for Raman scattering \cite{Deveraux2007} will hence seamlessly transfer to the Weyl effective theory, only with the additional appearance of the new SU(2) gauge field ${\bf A}^a$.

If we assume idealized circumstances in which electrons have an infinite lifetime and do not interact with one another, then the Raman scattering on Weyl electrons vanishes in the effective mass approximation. The linear part of the spectrum is inhibited already at the level of the vertex function, while the quadratic part introduced with $m<\infty$ in the $E_{{\bf k},\sigma}$ dispersion falls short of contributing the trace and momentum integral in (\ref{Chi1}). The latter feature is observed in ordinary metals as well, where a finite quasiparticle lifetime dominates the Raman scattering rate.

\subsection{Weyl quasiparticles with infinite lifetime}

Here we embark on the derivation of the Raman scattering vertex beyond the effective mass approximation. The scattering process is schematically illustrated in Fig.\ref{RamanWeyl}. We will utilize the bare Green's functions (\ref{Green1}) of Weyl electrons in (\ref{Chi1}), and conveniently work in the representation that diagonalizes the Weyl Hamiltonian $H_{\bf k}$. The sum over intermediate quasiparticle states $|\gamma\rangle$ in (\ref{Gamma1}) contains low-energy states $|\gamma\rangle$ which belong to the nodal spectrum. The remaining part of that sum samples $|\gamma\rangle$ from the bands at higher energies where the conditions for the effective mass approximation usually hold. Since this approximation simply yields zero in the present problem, we may assume that the true contribution of high-energy intermediate states is small and negligible next to the low-energy part with $|\gamma\rangle$ from the nodal spectrum:
\begin{widetext}
\begin{eqnarray}\label{Gamma2}
\gamma_{{\bf k}\sigma,{\bf k}'\sigma'}^{\textrm{i,s}} &\approx& \hat{{\bf e}}_{\textrm{i}}^{\phantom{x}}\hat{{\bf e}}_{\textrm{s}}^{\phantom{x}}\langle{\bf k},\sigma|e^{i{\bf q}{\bf r}}|{\bf k}',\sigma'\rangle+\frac{1}{m}\sum_{{\bf k}''\sigma''}\Biggl\lbrack\frac{\langle{\bf k},\sigma|e^{-i{\bf q}_{\textrm{s}}{\bf r}}\hat{{\bf e}}_{\textrm{s}}(-i\boldsymbol{\nabla}+mv\boldsymbol{\sigma})|{\bf k}'',\sigma''\rangle\,\langle{\bf k}'',\sigma''|e^{i{\bf q}_{\textrm{i}}{\bf r}}\hat{{\bf e}}_{\textrm{i}}(-i\boldsymbol{\nabla}+mv\boldsymbol{\sigma})|{\bf k}',\sigma'\rangle}{\frac{k'^{2}}{2m}+\sigma' vk'-\frac{k''^{2}}{2m}-\sigma'' vk''+\omega_{\textrm{i}}} \nonumber \\
&& \quad+\frac{\langle{\bf k},\sigma|e^{i{\bf q}_{\textrm{i}}{\bf r}}\hat{{\bf e}}_{\textrm{i}}(-i\boldsymbol{\nabla}+mv\boldsymbol{\sigma})|{\bf k}'',\sigma''\rangle\,\langle{\bf k}'',\sigma''|e^{-i{\bf q}_{\textrm{s}}{\bf r}}\hat{{\bf e}}_{\textrm{s}}(-i\boldsymbol{\nabla}+mv\boldsymbol{\sigma})|{\bf k}',\sigma'\rangle}{\frac{k'^{2}}{2m}+\sigma' vk'-\frac{k''^{2}}{2m}-\sigma'' vk''-\omega_{\textrm{s}}}\Biggr\rbrack \ .
\end{eqnarray}
\end{widetext}
The quasiparticle states have been relabeled using Weyl electron quantum numbers $\bf k$ and $\sigma=\pm 1$ as $|\alpha\rangle = |{\bf k},\sigma\rangle$, $|\beta\rangle = |{\bf k}',\sigma'\rangle$, $|\gamma\rangle = |{\bf k}'',\sigma''\rangle$. The relevant matrix elements involving Weyl states have the form
\begin{eqnarray}\label{GammaMatrixElements}
\langle{\bf k},\sigma|e^{i{\bf q}{\bf r}}|{\bf k}',\sigma'\rangle &=& \langle\sigma\hat{{\bf k}}|\sigma'\hat{{\bf k}}'\rangle\delta_{{\bf k},{\bf k}'+{\bf q}}^{\phantom{*}} \\
\langle{\bf k}'',\sigma''|e^{i{\bf q}_{i}{\bf r}}(-i\boldsymbol{\nabla})|{\bf k}',\sigma'\rangle &=& {\bf k}'\langle\sigma''\hat{{\bf k}}''|\sigma'\hat{{\bf k}}'\rangle  \delta_{{\bf k}'',{\bf k}'+{\bf q}_{\textrm{i}}}^{\phantom{*}} \nonumber \\
\langle{\bf k}'',\sigma''|e^{i{\bf q}_{i}{\bf r}}(mv\boldsymbol{\sigma})|{\bf k}',\sigma'\rangle &=& mv \langle\sigma''\hat{{\bf k}}''|\boldsymbol{\sigma}|\sigma'\hat{{\bf k}}'\rangle \delta_{{\bf k}'',{\bf k}'+{\bf q}_{\textrm{i}}}^{\phantom{*}} \ . \nonumber
\end{eqnarray}
Note that momentum conservation ultimately imposes ${\bf k}={\bf k}'+{\bf q}$, with ${\bf q} = {\bf q}_{\textrm{i}}-{\bf q}_{\textrm{s}}$, in all terms of the Raman vertex. The remaining state overlaps are between the $S=\frac{1}{2}$ coherent-state spinors $|\sigma{\bf k}\rangle$ whose spins point in the indicated directions $\sigma{\bf k}$. Since ${\bf k}$, ${\bf k}'$ and ${\bf k}''$ differ by very small wavevectors $\mathcal{O}({\bf q})$, we can immediately anticipate
\begin{eqnarray}
\langle\sigma{\bf k}|\sigma'{\bf k}'\rangle &\approx& \delta_{\sigma\sigma'}+\mathcal{O}({\bf q}) \nonumber \\
\langle\sigma\hat{{\bf k}}|\boldsymbol{\sigma}|\sigma'\hat{{\bf k}}'\rangle &\approx& \sigma\hat{{\bf k}}\,\delta_{\sigma\sigma'}+(\hat{{\bf k}}\times\boldsymbol{\lambda})(1-\delta_{\sigma\sigma'})+\mathcal{O}({\bf q}) \ . \nonumber
\end{eqnarray}
When the quasiparticles have infinite lifetime, only the zero-momentum particle-hole excitations of Weyl electrons can contribute to the Raman scattering at zero temperature, so the ``large'' $\sigma=\sigma'$ intraband part of the spinor overlap $\langle\sigma{\bf k}|\sigma'{\bf k}'\rangle$ will have no effect. Nevertheless, the presence of spin-orbit coupling will provide an $\mathcal{O}(1)$ contribution to the Raman vertex even in the ${\bf q}\to 0$ limit. Neglecting ${\bf q}_{\textrm{i,s}}$ amounts to equating ${\bf k}'$, ${\bf k}''$ and ${\bf k}$ in (\ref{Gamma2}):
\begin{eqnarray}
&& \gamma_{{\bf k}\sigma,{\bf k}'\sigma'}^{\textrm{i,s}}\Bigr\vert_{\sigma'=-\sigma} \approx v\delta_{{\bf k},{\bf k}'+{\bf q}}^{\phantom{*}} \\
&& \quad \times\Biggl\lbrack\left(\frac{ k+\sigma mv}{\omega_{\textrm{i}}-2\sigma vk}-\frac{ k-\sigma mv}{\omega_{\textrm{s}}}\right)(\hat{{\bf e}}_{\textrm{s}}\hat{{\bf k}})\Bigl(\hat{{\bf e}}_{\textrm{i}}\langle\sigma\hat{{\bf k}}|\boldsymbol{\sigma}|\sigma'\hat{{\bf k}}\rangle\Bigr) \nonumber \\
&& \quad +\left(\frac{ k-\sigma mv}{\omega_{\textrm{i}}}-\frac{ k+\sigma mv}{\omega_{\textrm{s}}+2\sigma vk}\right)(\hat{{\bf e}}_{\textrm{i}}\hat{{\bf k}})\Bigl(\hat{{\bf e}}_{\textrm{s}}\langle\sigma\hat{{\bf k}}|\boldsymbol{\sigma}|\sigma'\hat{{\bf k}}\rangle\Bigr)\Biggr\rbrack \nonumber 
\end{eqnarray}
The conservation of energy $\Omega = \omega_{\textrm{i}}-\omega_{\textrm{s}} = 2\sigma vk$ is formally enforced in (\ref{Rate1}) and (\ref{Chi1}) when the quasiparticles have infinite lifetime. Then, the vertex function simplifies:
\begin{eqnarray}\label{Gamma3}
&& \gamma_{{\bf k}\sigma,{\bf k}'\sigma'}^{\textrm{i,s}}\Bigr\vert_{\sigma'=-\sigma} \approx 2\sigma mv^{2}\delta_{{\bf k},{\bf k}'+{\bf q}}^{\phantom{*}} \\
&& \quad \times\Biggl\lbrack\frac{(\hat{{\bf e}}_{\textrm{s}}\hat{{\bf k}})\Bigl(\hat{{\bf e}}_{\textrm{i}}\langle\sigma\hat{{\bf k}}|\boldsymbol{\sigma}|\sigma'\hat{{\bf k}}\rangle\Bigr)}{\omega_{\textrm{s}}}-\frac{(\hat{{\bf e}}_{\textrm{i}}\hat{{\bf k}})\Bigl(\hat{{\bf e}}_{\textrm{s}}\langle\sigma\hat{{\bf k}}|\boldsymbol{\sigma}|\sigma'\hat{{\bf k}}\rangle\Bigr)}{\omega_{\textrm{i}}}\Biggr\rbrack \nonumber
\end{eqnarray}
The same simplification can be used more generally in the limit $\Omega, vk\ll\omega_{\textrm{i}}, \omega_{\textrm{s}}$.

We can now substitute the vertex function in (\ref{Chi1}) at ${\bf q}\to 0$ and evaluate the trace in the representation that diagonalizes the Weyl Hamiltonian:
\begin{eqnarray}\label{Chi2}
&& \chi({\bf q}\to0,\Omega) = -i \int\frac{d^3 k}{(2\pi)^3}\frac{d\omega}{2\pi} \sum_{\sigma\sigma'} \gamma^{\textrm{i,s}}_{{\bf k}\sigma,{\bf k}'\sigma'}\gamma^{\textrm{s,i}}_{{\bf k}'\sigma',{\bf k}\sigma} \nonumber \\
&& ~~\times \frac{1}{\omega - E_{{\bf k}'\sigma'} + i\textrm{sign}(E_{{\bf k}'\sigma'})}\, \frac{1}{\Omega+\omega - E_{{\bf k}\sigma} + i\textrm{sign}(E_{{\bf k}\sigma})} \nonumber
\end{eqnarray}
Only the terms with $\sigma'=-\sigma$ survive the frequency integration when ${\bf q}\to 0$, and hence (${\bf k}'\to{\bf k}$):
\begin{eqnarray}\label{Chi2b}
&& \chi({\bf q}\to0,\Omega) = \int\frac{d^{3}k}{(2\pi)^{3}}\sum_{\sigma}\gamma_{\sigma,-\sigma}^{\textrm{i,s}}({\bf k})\,\gamma_{-\sigma,\sigma}^{s,i}({\bf k}) \\
&& \quad\times \frac{\theta(-E_{{\bf k},-\sigma})-\theta(-E_{{\bf k},\sigma})}{\Omega-2\sigma vk-i0^{+}\textrm{sign}(E_{{\bf k},-\sigma})+i0^{+}\textrm{sign}(E_{{\bf k},\sigma})} \nonumber \ .
\end{eqnarray}
The product of the vertex functions is real, so that the Raman scattering ``rate'' (\ref{Rate1}) becomes:
\begin{eqnarray}\label{Rate2}
R &=& \frac{1}{2 v}\int\frac{d^{3}k}{(2\pi)^{3}}\,\gamma_{\sigma,-\sigma}^{\textrm{i,s}}({\bf k})\,\gamma_{-\sigma,\sigma}^{\textrm{s,i}}({\bf k}) \\
&& ~~\times\;\theta\left(\frac{|\Omega|}{2}-\left\vert \mu-\frac{\Omega^{2}}{8mv^{2}}\right\vert \right)\,\delta\left(k-\frac{|\Omega|}{2 v}\right)\,\delta_{\sigma,\textrm{sign}(\Omega)} \nonumber \\
&=& \frac{m^{2}v\Omega^{2}}{2(2\pi)^{3}}\;\theta\left(\frac{|\Omega|}{2}-\left\vert \mu-\frac{\Omega^{2}}{8mv^{2}}\right\vert \right)\times I(\hat{{\bf e}}_{\textrm{s}},\hat{{\bf e}}_{\textrm{i}}) \ . \nonumber
\end{eqnarray}
This already reveals the full frequency dependence of the Raman scattering cross-section. We observe $R\sim \Omega^2$ above a threshold Raman shift frequency, $|\Omega|>2|\mu|$  in the case of a perfectly linear Weyl spectrum ($m\to\infty$). It turns out that such a quadratic frequency dependence is not easy to obtain by other mechanisms, so it can serve as a good indicator of Weyl electrons. The overall factor $m^2$ naively looks problematic in the $m\to\infty$ limit, but it exactly cancels out with the factor of $m^{-2}$ in the scattering cross-section (\ref{CrossSection1}). Ultimately, the scattering cross-section is well-defined and mass-independent for perfectly relativistic Weyl electrons.

The dependence of the Raman scattering cross-section on the polarization of light is universal in our idealized model and contained in:
\begin{eqnarray}
&& I(\hat{{\bf e}}_{\textrm{s}},\hat{{\bf e}}_{\textrm{i}}) = \int\limits_{0}^{2\pi}d\phi\int\limits_{0}^{\pi}d\theta\,\sin\theta \\
&& ~~\times \left\vert \frac{(\hat{{\bf e}}_{s}\hat{{\bf k}})\Bigl(\hat{{\bf e}}_{i}\langle\hat{{\bf k}}|\boldsymbol{\sigma}|-\!\hat{{\bf k}}\rangle\Bigr)}{\omega_{\textrm{s}}}-\frac{(\hat{{\bf e}}_{i}\hat{{\bf k}})\Bigl(\hat{{\bf e}}_{s}\langle\hat{{\bf k}}|\boldsymbol{\sigma}|-\!\hat{{\bf k}}\rangle\Bigr)}{\omega_{\textrm{i}}}\right\vert ^{2} \ . \nonumber
\end{eqnarray}
The angles $\theta,\phi$ refer to the orientation of the wavevector $\hat{{\bf k}}=\hat{{\bf x}}\sin\theta\cos\phi+\hat{{\bf y}}\sin\theta\sin\phi+\hat{{\bf z}}\cos\theta$ in the integral from (\ref{Rate2}). We use the following representation of the spin coherent states, which properly regularizes the vorticity of the angle $\phi$,
\begin{equation}\label{SpinCoherentStates}
|+\hat{\bf k}\rangle = \left(\begin{array}{c}
\cos\left(\frac{\theta}{2}\right)\\
e^{i\phi}\sin\left(\frac{\theta}{2}\right)
\end{array}\right) \quad,\quad
|-\hat{\bf k}\rangle = \left(\begin{array}{c}
e^{-i\phi}\sin\left(\frac{\theta}{2}\right)\\
-\cos\left(\frac{\theta}{2}\right)
\end{array}\right)
\end{equation}
to finally obtain
\begin{eqnarray}\label{PolarI1}
I(\hat{{\bf e}}_{\textrm{s}},\hat{{\bf e}}_{\textrm{i}}) &=& \frac{16\pi}{15}\left(\frac{1}{\omega_{\textrm{s}}^{2}}+\frac{1}{\omega_{\textrm{i}}^{2}}\right)\left(1-\frac{(\hat{{\bf e}}_{\textrm{i}}\hat{{\bf e}}_{\textrm{s}})^{2}}{2}\right) \nonumber \\
&& +\frac{8\pi}{15}\frac{1}{\omega_{\textrm{i}}^{\phantom{x}}\omega_{\textrm{s}}^{\phantom{x}}}\Bigl(1-3(\hat{{\bf e}}_{\textrm{i}}\hat{{\bf e}}_{\textrm{s}})^{2}\Bigr) \ .
\end{eqnarray}
This polarization-dependence is universal but realistically tainted in a finite frequency range as we discuss in the next section. If one aligns the $z$-axis with the incident light polarization $\hat{\bf e}_{\textrm{i}}$, and the outgoing light polarization $\hat{\bf e}_{\textrm{s}}$ points in the $(\theta,\phi)$ direction in this spherical coordinate system, then $I(\hat{{\bf e}}_{\textrm{s}},\hat{{\bf e}}_{\textrm{i}}) \equiv I(\theta,\phi)$ is a specific linear combination of $s$ and $d_{z^2}$ spherical harmonics given by (\ref{PolarI1}).

\subsection{Tilted Weyl spectrum}

A Weyl semimetal generally has a number of Weyl nodes scattered throughout the first Brillouin zone, and lattice symmetries permit the Weyl spectrum of any particular node to have a tilted form
\begin{equation}\label{WeylEnergy2}
E_{{\bf k}\sigma} = \frac{k^{2}}{2m} - {\bf u}{\bf k} + \sigma vk - \mu \ .
\end{equation}
The tilt velocity ${\bf u}$ is related to the formal U(1) gauge shift discussed after Eq.\ref{WeylEnergy1} by ${\bf a} = m {\bf u}$. Different tilt velocities are required on different nodes by point-group and time-reversal symmetries, so the polarization dependence of the Raman response may acquire a less universal form which reflects the lattice symmetries. We will assume $v>u$ for the type-I Weyl spectrum, and focus on the relativistic $m\to\infty$ limit where $\sigma=\textrm{sign}(\mu)$ is ensured at the Fermi surface, see Fig.\ref{RamanWeyl}(b).

A finite chemical potential $\mu$ shapes a Weyl Fermi surface which is not centered any more at the origin of momentum space. The Fermi wavevector of the Weyl Fermi surface measured from the origin
\begin{equation}
k_{F}(\hat{\bf k}) \xrightarrow{m\to\infty} \frac{|\mu|}{v-({\bf u}\hat{{\bf k}})\textrm{sign}(\mu)}
\end{equation}
now depends on the wavevector direction $\hat{\bf k}$. Fortunately, the interband Raman vertex functions $\gamma_{\bf k}$ are unaffected by ${\bf u}\neq0$ and still given by (\ref{Gamma3}) in the limit ${\bf q}\to0$, $\Omega \ll \omega_{\textrm{i}},\omega_{\textrm{s}}$ (the plain momentum shift $m{\bf u}$ disappears along with the momentum ${\bf k}$ from the $\gamma_{\bf k}$ expression).

\begin{figure}
\includegraphics[width=3.2in]{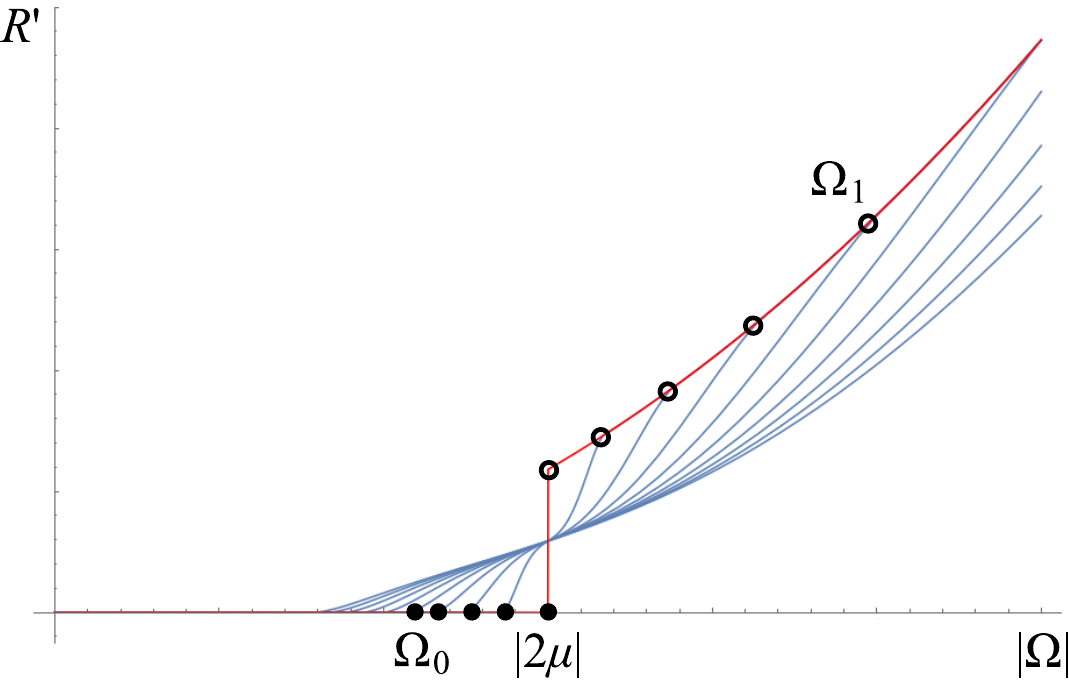}
\caption{\label{Wtilted}Interband Raman scattering rate from quasiparticles with a tilted Weyl spectrum and infinite lifetime. The plots are parametrized by the tilt $u/v$, starting from zero (red curve) and growing in increments $0.1$ until $0.9$ (blue curves). The unique threshold frequency at $u/v=0$ splits into a lower threshold frequency $\Omega_0$ (solid circles) below which the Raman response vanishes, and an upper threshold frequency $\Omega_1$ (open circles) above which the Raman response falls back to the most universal form obtained in the absence of tilt. The polarization vectors $\hat{\bf e}_{\textrm{i}}(\theta_{\textrm{i}},\phi_{\textrm{i}})$ and $\hat{\bf e}_{\textrm{s}}(\theta_{\textrm{s}},\phi_{\textrm{s}})$ for this calculation were $\theta_{\textrm{i}}=0.14$, $\phi_{\textrm{i}}=2.87$, and $\theta_{\textrm{s}}=0.40$, $\phi_{\textrm{s}}=2.34$ respectively.}
\end{figure}

The interband Raman response of the Weyl quasiparticles with infinite lifetime is obtained from (\ref{Chi2b}). Taking advantage of the fact that the vertex functions are approximately independent of $k=|{\bf k}|$, we have
\begin{eqnarray}
&& R'(\Omega) = \int\frac{d^{3}k}{(2\pi)^{3}}\sum_{\sigma} \theta\left(k-k_{F}(\hat{{\bf k}})\right)\,\delta(\Omega-2\sigma vk) \nonumber \\
&& \qquad\qquad\times \gamma_{\sigma,-\sigma}^{\textrm{i,s}}({\bf k})\,\gamma_{-\sigma,\sigma}^{\textrm{s,i}}({\bf k}) \nonumber \\
&&\quad = \frac{1}{(2\pi)^{3}}\frac{1}{2v}\left(\frac{|\Omega|}{2v}\right)^{2}\int d^{2}\hat{{\bf k}}\:\theta\left(\frac{|\Omega|}{2}-vk_{F}(\hat{{\bf k}})\right) \nonumber \\
&& \qquad\qquad\times \gamma_{\sigma,-\sigma}^{\textrm{i,s}}(\hat{\bf k})\,\gamma_{-\sigma,\sigma}^{\textrm{s,i}}(\hat{\bf k})\,\Bigr\vert_{\sigma=\textrm{sign}(\Omega)} \ .
\end{eqnarray}
The remaining integral over $\hat{\bf k}$ is frequency dependent and affected by two threshold frequencies:
\begin{eqnarray}
\Omega_{0} &=& \min\left(2|\mu|+\frac{2({\bf u}\hat{{\bf k}})\mu}{v-({\bf u}\hat{{\bf k}})\textrm{sign}(\mu)}\right) = \frac{|2\mu|}{1+\frac{u}{v}} \\
\Omega_{1} &=& \max\left(2|\mu|+\frac{2({\bf u}\hat{{\bf k}})\mu}{v-({\bf u}\hat{{\bf k}})\textrm{sign}(\mu)}\right) = \frac{|2\mu|}{1-\frac{u}{v}} \ . \nonumber
\end{eqnarray}
The scattering rate vanishes when $|\Omega|<\Omega_0$, and falls back to the $u=0$ form when $|\Omega|>\Omega_1$. The Weyl spectrum tilt $u$ replaces the sudden jump of $R(\Omega)$ by a gradual rise from zero at $|\Omega|=\Omega_0$ to $R\sim\Omega^2$ at $|\Omega|>\Omega_1$, as shown in Fig.\ref{Wtilted}. This makes it more difficult to experimentally determine the chemical potential, and we will show in the next section that the thresholds are blurred even further due to the finite quasiparticle lifetime. Nevertheless, in idealized circumstances (infinite lifetime, all Weyl nodes having the same $u/v$ and $|\mu|$, no additional sources of Raman signal at higher frequencies), one would be able to extract the ratio $u/v$ and the chemical potential by measuring both threshold frequencies:
\begin{equation}
\frac{u}{v}=\frac{\Omega_{1}-\Omega_{0}}{\Omega_{1}+\Omega_{0}}\quad,\quad|\mu|=\frac{\Omega_{0}\Omega_{1}}{\Omega_{1}+\Omega_{0}} \ .
\end{equation}
Detecting the upper threshold $\Omega_1$ would be particularly challenging. It is necessary either to confirm a ``pure'' $R(\Omega)\propto\Omega^2$ frequency dependence above $|\Omega|>\Omega_1$, or carefully examine the light polarization dependence of the Raman response. At frequencies $|\Omega|>\Omega_1$, the polarization dependence is frequency-independent and universally given by (\ref{PolarI1}). In the intermediate frequency range $\Omega_0<|\Omega|<\Omega_1$, the polarization dependence is modified and acquires a frequency dependence from
\begin{eqnarray}
&& I(\hat{{\bf e}}_{\textrm{s}},\hat{{\bf e}}_{\textrm{i}},|\Omega|) = \int\limits_{0}^{2\pi}d\phi\int\limits_{0}^{\pi}d\theta\,\sin\theta \; \theta\left(1-\frac{u}{v}\cos\theta-\frac{2|\mu|}{|\Omega|}\right) \nonumber \\
&& ~~\times \left\vert \frac{(\hat{{\bf e}}_{s}\hat{{\bf k}})\Bigl(\hat{{\bf e}}_{i}\langle\hat{{\bf k}}|\boldsymbol{\sigma}|-\!\hat{{\bf k}}\rangle\Bigr)}{\omega_{\textrm{s}}}-\frac{(\hat{{\bf e}}_{i}\hat{{\bf k}})\Bigl(\hat{{\bf e}}_{s}\langle\hat{{\bf k}}|\boldsymbol{\sigma}|-\!\hat{{\bf k}}\rangle\Bigr)}{\omega_{\textrm{i}}}\right\vert ^{2} \ . \nonumber
\end{eqnarray}

\subsection{Weyl quasiparticles with a finite lifetime}

Interactions and disorder generally introduce a finite lifetime for the quasiparticles through the imaginary part of their self-energy correction. We will model this with a prototype Green's function in the basis of Weyl electron states
\begin{equation}\label{Green2}
G_\sigma({\bf k},\omega)=\frac{A_\sigma({\bf k},\omega)}{\omega-E_{{\bf k}\sigma}+i\Gamma \textrm{sign}(\omega)} \ .
\end{equation}
For simplicity, we will focus again on the non-tilted Weyl spectrum given by (\ref{WeylEnergy1}). Instead of calculating the self-energy correction $\Sigma({\bf k},\omega) = \delta E({\bf k},\omega) - i\Gamma({\bf k},\omega)$, we will make perhaps a simplistic assumption that its main effect of concern is a finite lifetime $\tau$ that the quasiparticles acquire. If this lifetime is not too short, then the concrete frequency and momentum dependence of $\Gamma({\bf k},\omega)$ is not crucial and may be replaced by a constant $\Gamma\propto\tau^{-1}$ multiplying $\textrm{sign}(\omega)$ which is required by the analytic properties of the Green's function. The real part of $\Sigma$ is to be absorbed into the renormalized Weyl electron energy $E_{{\bf k}\sigma}$. Assuming that the spectrum is not qualitatively changed by interactions or disorder, we will keep the expression (\ref{WeylEnergy1}) with understanding that its parameters are renormalized. Finally, in order to make analytical progress, we shall also neglect the dependence of the spectral weight $A_\sigma({\bf k},\omega)$ on the momentum and frequency. The final formulas will simply assume $A=1$, but a small reduction of the predicted scattering rates should be realistically expected.

We are now ready to embark on the calculation of (\ref{Chi1}) with the modelled Green's functions (\ref{Green2}):
\begin{eqnarray}\label{Chi3}
\chi({\bf q}\to0,\Omega) &=& -i \int\frac{d^3 k}{(2\pi)^3}\frac{d\omega}{2\pi} \sum_{\sigma\sigma'} \gamma^{\textrm{i,s}}_{{\bf k}\sigma,{\bf k}'\sigma'}\gamma^{\textrm{s,i}}_{{\bf k}'\sigma',{\bf k}\sigma} \nonumber \\
&& \times \, G_{\sigma'}({\bf k'},\omega) \, G_\sigma({\bf k},\Omega+\omega) \ .
\end{eqnarray}
We will separately consider interband ($\sigma'=-\sigma$) and intraband ($\sigma'=\sigma$) processes. We already calculated the vertex functions (\ref{Gamma3}) for the interband process; note that the Raman shift frequency $\Omega$ is not pinned to $2\sigma vk$ any more due to $\Gamma\neq 0$, but the expression (\ref{Gamma3}) still holds in the limit $\Omega, vk\ll\omega_{\textrm{i}}, \omega_{\textrm{s}}$, which we will adopt for simplicity. Then, the interband part of the Raman response function is:
\begin{equation}
\chi'({\bf q}\to0,\Omega) = -i\int\frac{d^{3}k}{(2\pi)^{3}}\sum_{\sigma}\gamma_{\sigma,-\sigma}^{\textrm{i,s}}({\bf k})\,\gamma_{-\sigma,\sigma}^{\textrm{s,i}}({\bf k})\, I_{\sigma}^{\phantom{x}}(\Omega) \ , \nonumber
\end{equation}
where
\begin{eqnarray}
I_{\sigma}(\Omega) &=& \int\frac{d\omega}{2\pi}\,\frac{1}{\omega-\frac{\Omega}{2}-E_{{\bf k},-\sigma}+i\Gamma\textrm{sign}\left(\omega-\frac{\Omega}{2}\right)} \quad \\
&& \qquad\times \frac{1}{\omega+\frac{\Omega}{2}-E_{{\bf k},\sigma}+i\Gamma\textrm{sign}\left(\omega+\frac{\Omega}{2}\right)} \ . \nonumber
\end{eqnarray}
It is helpful to observe that $I_{\sigma}(\Omega)=I_{-\sigma}(-\Omega)$ and
\begin{equation}
I_{\sigma}(\Omega)=I_{\sigma}(|\Omega|)\,\theta(\Omega)+I_{\sigma}(-|\Omega|)\,\theta(-\Omega)=I_{\sigma\,\textrm{sign}(\Omega)}(|\Omega|) \nonumber
\end{equation}
Together with the earlier finding that $\gamma_{\sigma,-\sigma}^{\textrm{i,s}}({\bf k})\,\gamma_{-\sigma,\sigma}^{\textrm{s,i}}({\bf k})$ is $\sigma$-independent, this will ensure that the Raman scattering rate depends only on $|\Omega|$. Integrating out the frequency $\omega$ can now be easily performed directly, without the use of Cauchy's residue theorem. After this, the imaginary part of $\chi'$ yields the interband contribution to the Raman scattering rate (\ref{Rate1}):
\begin{eqnarray}\label{Rate3}
R'(\Omega) &=& \frac{1}{\pi}\int\frac{d^{3}k}{(2\pi)^{3}}\sum_{\sigma}\gamma_{\sigma,-\sigma}^{\textrm{i,s}}({\bf k})\,\gamma_{-\sigma,\sigma}^{\textrm{s,i}}({\bf k})\,\textrm{Re}\left\lbrace I_{\sigma}^{\phantom{x}}(\Omega)\right\rbrace \nonumber \\
&=& \frac{(mv^{2})^{2}}{2\pi^{4}}I(\hat{{\bf e}}_{\textrm{s}},\hat{{\bf e}}_{\textrm{i}})\int\limits_{0}^{\infty}dk\,k^{2}\sum_{\sigma}\textrm{Re}\left\lbrace I_{\sigma}(|\Omega|)\right\rbrace \quad
\end{eqnarray}
with $I(\hat{{\bf e}}_{\textrm{s}},\hat{{\bf e}}_{\textrm{i}})$ given by (\ref{PolarI1}) and
\begin{widetext}
\begin{eqnarray}\label{ReIsigma}
&& 2\pi\,\textrm{Re}\left\lbrace I_{\sigma}(|\Omega|)\right\rbrace = \frac{1}{2}\left(\frac{|\Omega|-\Delta E_{{\bf k}\sigma}}{(|\Omega|-\Delta E_{{\bf k}\sigma})^{2}+4\Gamma^{2}}-\frac{1}{|\Omega|-\Delta E_{{\bf k}\sigma}}\right)\left\lbrack \log\left(\frac{E_{{\bf k},\sigma}^{2}+\Gamma^{2}}{(E_{{\bf k},-\sigma}+|\Omega|)^{2}+\Gamma^{2}}\right)+\log\left(\frac{E_{{\bf k},-\sigma}^{2}+\Gamma^{2}}{(E_{{\bf k},\sigma}-|\Omega|)^{2}+\Gamma^{2}}\right)\right\rbrack \nonumber \\
&&\quad+\frac{2\Gamma}{(|\Omega|-\Delta E_{{\bf k}\sigma})^{2}+4\Gamma^{2}}\,\Biggl\lbrack\arctan\left(\frac{E_{{\bf k},-\sigma}+|\Omega|}{\Gamma}\right)-\arctan\left(\frac{E_{{\bf k},\sigma}-|\Omega|}{\Gamma}\right)+\arctan\left(\frac{E_{{\bf k},\sigma}}{\Gamma}\right)-\arctan\left(\frac{E_{{\bf k},-\sigma}}{\Gamma}\right)\Biggr\rbrack \ .
\end{eqnarray}
\end{widetext}
Here, $\Delta E_{{\bf k}\sigma} = E_{{\bf k},\sigma} - E_{{\bf k},-\sigma}$ in general, and $\Delta E_{{\bf k}\sigma} = 2\sigma vk$ for the Weyl spectrum. The last one-dimensional integration over $k$ in (\ref{Rate3}) has to be done numerically, but it is straight-forward.

\begin{figure}
\includegraphics[width=3.2in]{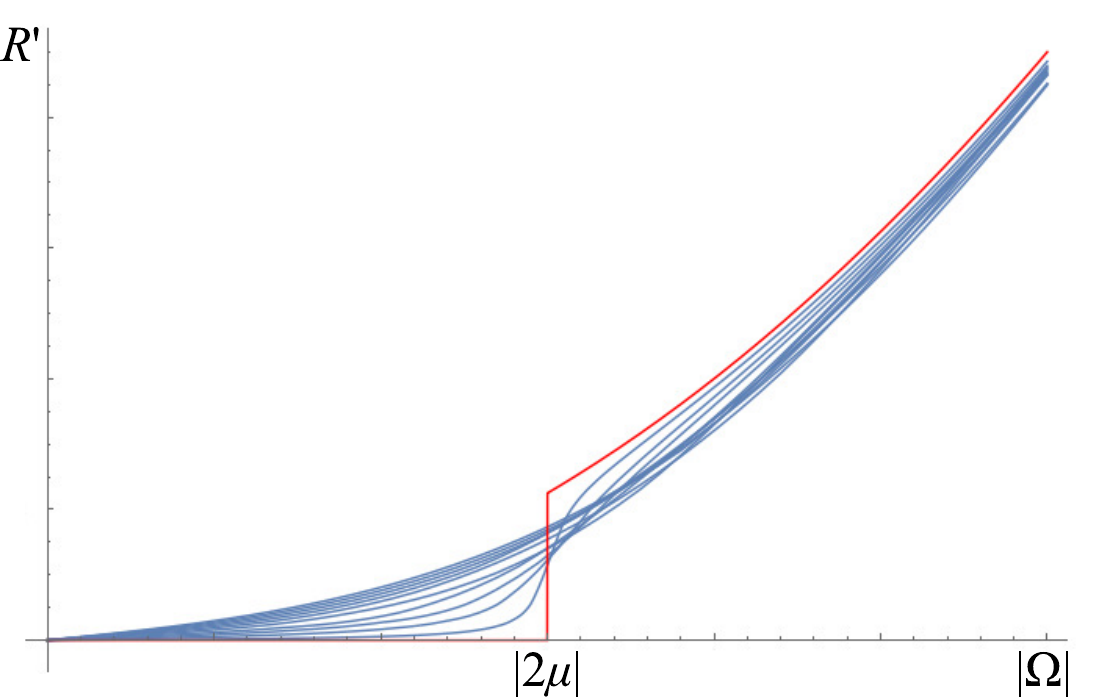}
\caption{\label{Winterband}Interband Raman scattering rate from spherically-symmetric Weyl quasiparticles parametrized by the quasiparticle lifetime $\tau$. The Raman shift frequency $\Omega$ and the scattering rate $R'$ are shown in arbitrary units. The red curve corresponds to infinite lifetime -- scattering occurs only above a threshold frequency $\Omega>|2\mu|$, which is determined by the chemical potential $\mu$ relative to the node energy, and proceeds as $R'\propto\Omega^2$. The blue curves illustrate the evolution of Raman scattering as the quasiparticle ``decay rate'' $\Gamma\propto\tau^{-1}$ increases from $0.05$ in steps $0.1$ (expressed with the same energy units as $\Omega$). The scattering below the ``threshold'' quickly fills up with a linear-looking dependence on $\Omega$.}
\end{figure}

The effect of the finite Weyl quasiparticle lifetime on the interband Raman scattering is shown in Fig.\ref{Winterband}. The main consequence of the processes that decay the Weyl quasiparticles is a rapid emergence of scattering at low Raman shift frequencies, below the previous threshold $\Omega=|2\mu|$. The Raman scattering rate becomes approximately a linear function of $\Omega$ at low frequencies and crosses over to the earlier result $R\sim\Omega^2$ shortly beyond the ``threshold''. The sudden onset of scattering is easily washed out when the quasiparticles have a finite lifetime, and hence likely hard to observe in experiments.

Next, we turn our attention to the intraband scattering and calculate the $\sigma'=\sigma$ part of (\ref{Chi3}):
\begin{eqnarray}
\chi''({\bf q}\to0,\Omega) &=& -i \int\frac{d^3 k}{(2\pi)^3}\frac{d\omega}{2\pi} \sum_{\sigma} \gamma^{\textrm{i,s}}_{{\bf k}\sigma,{\bf k}'\sigma}\gamma^{\textrm{s,i}}_{{\bf k}'\sigma,{\bf k}\sigma} \nonumber \\
&& \times \, G_{\sigma}({\bf k'},\omega) \, G_\sigma({\bf k},\Omega+\omega) \ .
\end{eqnarray}
A direct integration of $\omega$ is straight-forward with $\Gamma\neq0$ in the Green's functions, and it quickly becomes apparent that the result is not zero at finite $\Omega$ as before when $\Gamma$ was infinitesimal. After realizing that $\chi''$ depends only on $|\Omega|$ and taking the imaginary part, we get the intraband contribution to the Raman scattering ``rate'':
\begin{eqnarray}\label{Rate4}
&& R''(\Omega) = \frac{1}{2\pi^{2}}\,\frac{2\Gamma}{|\Omega|^{2}+4\Gamma^{2}}\int\frac{d^{3}k}{(2\pi)^{3}}\sum_{\sigma}\gamma_{\sigma\sigma}^{\textrm{i,s}}({\bf k})\,\gamma_{\sigma\sigma}^{\textrm{s,i}}({\bf k}) \nonumber \\
&& \quad\times \Biggl\lbrace -\frac{2\Gamma}{|\Omega|}\,\log\left(\frac{E_{{\bf k}\sigma}^{2}+\Gamma^{2}}{\sqrt{(E_{{\bf k}\sigma}^{2}-|\Omega|^{2}+\Gamma^{2})^{2}+4\Gamma^{2}|\text{\ensuremath{\Omega}}|^{2}}}\right) \nonumber \\
&& \qquad~ +\arctan\left(\frac{|\Omega|+E_{{\bf k}\sigma}^{\phantom{x}}}{\Gamma}\right)+\arctan\left(\frac{|\Omega|-E_{{\bf k}\sigma}^{\phantom{x}}}{\Gamma}\right)\Biggr\rbrace \nonumber
\end{eqnarray}
The needed product of intraband vertex functions can be again derived from (\ref{Gamma2}) and (\ref{GammaMatrixElements}). The result is complicated unless we take the limit  $\Omega,vk\ll\omega_{\textrm{i}},\omega_{\textrm{s}},mv^{2}$ and drop the terms of order $\mathcal{O}({\bf q})$:
\begin{equation}
\gamma_{\sigma\sigma}^{\textrm{i,s}}({\bf k})\,\gamma_{\sigma\sigma}^{\textrm{s,i}}({\bf k}) \approx \delta_{{\bf k},{\bf k}'+{\bf q}}^{\phantom{*}}\left\lbrack (\hat{{\bf e}}_{\textrm{i}}^{\phantom{x}}\hat{{\bf e}}_{\textrm{s}}^{\phantom{x}})^{2}+\left(\textrm{Im}\lbrace X_{\sigma}^{\textrm{i,s}}(\hat{{\bf k}})\rbrace\right)^{2}\right\rbrack \ . \nonumber
\end{equation}
When the function
\begin{eqnarray}
X_{\sigma}^{\textrm{i,s}}(\hat{{\bf k}}) &=& mv^{2}\,\frac{\omega_{\textrm{s}}+\omega_{\textrm{i}}}{\omega_{\textrm{i}}\omega_{\textrm{s}}} \nonumber \\
&& \times \Bigl(\hat{{\bf e}}_{\textrm{s}}\langle\sigma\hat{{\bf k}}|\boldsymbol{\sigma}|-\sigma\hat{{\bf k}}\rangle\Bigr)\Bigl(\hat{{\bf e}}_{\textrm{i}}\langle-\sigma\hat{{\bf k}}|\boldsymbol{\sigma}|\sigma\hat{{\bf k}}\rangle\Bigr) \ . \nonumber
\end{eqnarray}
is calculated in the representation (\ref{SpinCoherentStates}), we find
\begin{eqnarray}
\gamma_{\sigma\sigma}^{\textrm{i,s}}({\bf k})\,\gamma_{\sigma\sigma}^{\textrm{s,i}}({\bf k}) &\approx& \delta_{{\bf k},{\bf k}'+{\bf q}}^{\phantom{*}}\Biggr\lbrack (\hat{{\bf e}}_{\textrm{i}}^{\phantom{x}}\hat{{\bf e}}_{\textrm{s}}^{\phantom{x}})^{2} \\
&& +\left(\frac{mv^{2}(\omega_{\textrm{s}}+\omega_{\textrm{i}})}{\omega_{\textrm{i}}\omega_{\textrm{s}}}\right)^{2}\left(\hat{{\bf k}}(\hat{{\bf e}}_{\textrm{s}}\times\hat{{\bf e}}_{\textrm{i}})\right)^{2}\Biggr\rbrack \nonumber \ ,
\end{eqnarray}
and then substituting everything into (\ref{Rate4}) allows us to analytically carry out the integration over the directions of ${\bf k}$:
\begin{eqnarray}\label{Rate4b}
R''(\Omega) &=& \frac{1}{4\pi^{4}}\,\frac{2\Gamma}{|\Omega|^{2}+4\Gamma^{2}} \left(\frac{\Gamma}{v}\right)^3 I''\left(\frac{|\Omega|}{\Gamma}, \frac{\mu}{\Gamma}\right) \\
&& \times \left\lbrack (\hat{{\bf e}}_{\textrm{i}}^{\phantom{x}}\hat{{\bf e}}_{\textrm{s}}^{\phantom{x}})^{2}\!+\!\left(\frac{mv^{2}(\omega_{\textrm{s}}+\omega_{\textrm{i}})}{\omega_{\textrm{i}}\omega_{\textrm{s}}}\right)^{\!\!2}\frac{|\hat{{\bf e}}_{\textrm{s}}\times\hat{{\bf e}}_{\textrm{i}}|^{2}}{3}\right\rbrack \ . \nonumber
\end{eqnarray}
The remaining dimensionless integral
\begin{eqnarray}\label{Ipp}
I''(\phi,\nu) &=& \sum_{\sigma}\int\limits _{0}^{\infty}d\kappa\,\kappa^{2} \\
&& \times \Biggl\lbrace -\frac{2}{\phi}\,\log\left(\frac{\xi_{k\sigma}^{2}+1}{\sqrt{(\xi_{k\sigma}^{2}-\phi^{2}+1)^{2}+4\phi^{2}}}\right) \nonumber \\
&& \quad +\arctan\left(\phi+\xi_{k\sigma}\right)+\arctan\left(\phi-\xi_{k\sigma}\right)\Biggr\rbrace \nonumber 
\end{eqnarray}
with
\begin{equation}
\phi=\frac{|\Omega|}{\Gamma} \quad,\quad \nu=\frac{\mu}{\Gamma} \quad,\quad \xi_{k\sigma} = \frac{\Gamma\kappa^{2}}{2mv^{2}}+\sigma\kappa-\nu \nonumber
\end{equation}
has do be calculated numerically. Note that the factor of $1/\phi \propto 1/|\Omega|$ in the integral is tamed by the logarithm, and the $I''$ is well-behaved when $\Omega\to 0$. In fact, the frequency dependence of $I''$ is mild. For perfectly linearly-dispersing Weyl quasiparticles $m\to\infty$, the dominant behavior is $I''\approx 2.2\,\phi^{3}$ at $\mu=0$ and $I'' \propto \phi + \mathcal{O}(\phi^3)$ when $\mu\neq0$. Therefore, in the generic case of Weyl electrons with a Fermi surface, the intraband Raman scattering ``rate'' behaves approximately as
\begin{equation}
R''(\Omega) \sim \frac{\tau|\Omega|}{\tau^2\Omega^2+1} \ . 
\end{equation}
where $\tau\propto\Gamma^{-1}$ is the Weyl quasiparticle lifetime. Ordinary metals exhibit qualitatively the same Raman response, but unlike the conventional electrons, Weyl quasiparticles also produce the interband contribution which survives even in the $\Gamma\to 0$ limit.

\section{Quadratic band touching}\label{QBT}

Another type of nodal electron spectrum is quadratic band touching. Such a spectrum can temporarily arise when Weyl nodes coalesce, but it requires fine-tuning in general circumstances unless there is a symmetry to protect it \cite{Moon2013}. A temporary merger of two Weyl nodes with the same chirality would produce a topologically protected chiral charge-2 node with quadratic band touching. In contrast, a more likely merger of nodes with opposite chiralities eventually leads to a gap opening, with a Dirac spectrum or neutral quadratic band touching at the transition point depending on details such as the symmetry-imposed number of the merging Weyl nodes.

Here we analyze the Raman response of a Luttinger semimetal, where quadratic band touching is protected by the cubic and time reversal symmetries. This is relevant for pyrochlore iridates which can even exhibit phase transitions between Weyl and Luttinger semimetal states. The microscopically-motivated low-energy effective model for pyrochlore iridates is given by the Hamiltonian \cite{Moon2013}
\begin{equation}\label{HamiltonianLuttinger}
H_{\bf k}=\frac{k^{2}}{2M}+\frac{\frac{5}{4}k^{2}-({\bf k}{\bf J})^{2}}{2m'}-\frac{(k_{x}J_{x})^{2}\!+\!(k_{y}J_{y})^{2}\!+\!(k_{z}J_{z})^{2}}{2M_{\textrm{c}}}
\end{equation}
with three mass parameters $M, M_{\textrm{c}}, m'$. $J_i$ ($i\in\lbrace x,y,z \rbrace$) are the spin $S=\frac{3}{2}$ matrices in the standard four-dimensional representation. The spectrum
\begin{equation}\label{spectrumLuttinger}
E_{{\bf k}\sigma} = \frac{k^{2}}{2M'} + \sigma\sqrt{\left(\frac{k^{2}}{2m'}\right)^{2}+\frac{m'+2M_{\textrm{c}}}{4m'M_{\textrm{c}}^{2}}\,p_{\textrm{c}}({\bf k})} - \mu \ ,
\end{equation}
where
\begin{equation}
p_{\textrm{c}}({\bf k})=\sum_{i}k_{i}^{4}-\sum_{i>j}k_{i}^{2}k_{j}^{2}\quad,\quad M'=\frac{4MM_{\textrm{c}}}{4M_{\textrm{c}}-5M} \nonumber
\end{equation}
contains a conduction ($\sigma=1$) and a valence ($\sigma=-1$) band touching quadratically at ${\bf k}=0$ when $|M'|>m'$. Both bands are two-fold degenerate with respect to a ``flavor'' index $n\in\lbrace 1,2 \rbrace$. The dominant part of the Raman vertex function can be obtained using the effective mass approximation (\ref{Gamma1b})
\begin{eqnarray}\label{Gamma4}
\gamma_{{\bf k}}^{\phantom{x}} &\approx& m\,\hat{e}_{\textrm{i}}^{a}\hat{e}_{\textrm{s}}^{b}\frac{\partial^{2}H_{{\bf k}}}{\partial k^{a}\partial k^{b}} \\
&=& m\,\hat{e}_{\textrm{i}}^{a}\hat{e}_{\textrm{s}}^{b}\left(\frac{\delta_{ab}^{\phantom{x}}}{M}+\frac{\frac{5}{2}\delta_{ab}^{\phantom{x}}-\lbrace J_{a}^{\phantom{x}},J_{b}^{\phantom{x}}\rbrace}{2m'}-\frac{J_{a}^{2}\delta_{ab}^{\phantom{x}}}{M_{c}}\right) \ . \nonumber
\end{eqnarray}
Its matrix elements $\gamma^{\textrm{i,s}}_{\alpha\beta} = \langle\alpha|\gamma_{{\bf k}}|\beta\rangle$ in the basis of Hamiltonian eigenstates $|\alpha,\beta\rangle$ are generally non-zero for interband transitions. The states $|\alpha\rangle$ and $|\beta\rangle$ are to be taken at wavevectors separated by the photon's momentum transfer ${\bf q}\to 0$, but when one takes the realistic limit ${\bf q}\to 0$ and considers the states $|\alpha\rangle$ and $\beta\rangle$ at the same wavevector ${\bf k}$, one finds that $\gamma^{\textrm{i,s}}_{\alpha\beta}$ depend only on the direction of ${\bf k}$ and not on its magnitude $k=|{\bf k}|$. This is an important detail which will allow us to quickly find the frequency dependence of the Raman scattering rate. We will also find that going beyond the effective mass approximation introduces only a weak $k$-dependence into $\gamma_{{\bf k}}$, of the order of $\Omega/\omega_{\textrm{i,s}}$. 

Following the approach from the previous section, we model the Green's function of quasiparticles as (\ref{Green1}) or (\ref{Green2}) but with the spectrum (\ref{spectrumLuttinger}). The Raman response function (\ref{Chi1}) can be computed easily in the representation that diagonalizes the Hamiltonian. The most interesting features come from interband transitions, so we focus on those first. If the quasiparticles have infinite lifetime, then a formula analogous to (\ref{Chi2b}) obtains for interband processes:
\begin{eqnarray}\label{Chi4}
&& \chi'({\bf q}\to0,\Omega) = \sum_{nn'} \int\frac{d^{3}k}{(2\pi)^{3}}\sum_{\sigma}\gamma_{\sigma n,-\sigma n'}^{\textrm{i,s}}({\bf k})\,\gamma_{-\sigma n',\sigma n}^{\textrm{s,i}}({\bf k}) \nonumber \\
&& \quad\times \frac{\theta(-E_{{\bf k},-\sigma})-\theta(-E_{{\bf k},\sigma})}{\Omega\!-\!(E_{{\bf k},\sigma}\!-\!E_{{\bf k},-\sigma})\!-\!i0^{+}\textrm{sign}(E_{{\bf k},-\sigma})\!+\!i0^{+}\textrm{sign}(E_{{\bf k},\sigma})} \nonumber
\end{eqnarray}
Taking the imaginary part gives us the interband Raman scattering ``rate'' just as in the previous section:
\begin{eqnarray}\label{Rate5}
&& R'(\Omega) = \sum_{nn'} \int\frac{d^{2}\hat{\bf k}}{(2\pi)^{3}}\sum_{\sigma}\gamma_{\sigma n,-\sigma n'}^{\textrm{i,s}}(\hat{\bf k})\,\gamma_{-\sigma n',\sigma n}^{\textrm{s,i}}(\hat{\bf k}) \nonumber \\
&& \quad \times \int dk \,k^2\, \theta\Bigl( k-k_{\textrm{F}}(\hat{\bf k}) \Bigr)\, \delta\Bigl(\Omega\!-\!(E_{{\bf k},\sigma}\!-\!E_{{\bf k},-\sigma}) \Bigr) \ . \qquad
\end{eqnarray}
$k_{\textrm{F}}(\hat{\bf k})$ is the Fermi wavevector in the direction $\hat{\bf k}$ measured from the origin, for the band crossed by the chemical potential $\mu$. We took advantage of the fact that the Raman vertex functions in the effective mass approximation do not depend on the wavevector magnitude $k$.

If the detailed anisotropy of (\ref{spectrumLuttinger}) due to the cubic symmetry is neglected, then $k_{\textrm{F}}$ is independent of $\hat{\bf k}$ and $E_{{\bf k},\sigma}\!-\!E_{{\bf k},-\sigma} \approx \sigma k^2/ \widetilde{m}$ is approximately a quadratic function of $k$. A finite Raman response requires $|\Omega| = k^2/ \widetilde{m}$ to be larger than $k_{\textrm{F}}^2/ \widetilde{m}$, so integrating out $k$ immediately yields
\begin{equation}
R' (\Omega) \propto \sqrt{|\Omega|} \; \theta\Bigl( |\Omega|\!-\!(E_{{\bf k}_{\textrm{F}}+}\!-\!E_{{\bf k}_{\textrm{F}}-}) \Bigr) \ .
\end{equation}
If the spectrum has particle-hole symmetry at $\mu=0$, then the threshold frequency $E_{{\bf k}_{\textrm{F}}+}\!-\!E_{{\bf k}_{\textrm{F}}-}$ for Raman response is equal to $2|\mu|$.

When the detailed anisotropy of (\ref{spectrumLuttinger}) is not neglected, then we can still use the fact that the quasiparticle energy in the present model has the form $E_{{\bf k}\sigma} = k^2/2\widetilde{m}_{\hat{\bf k}\sigma}$ involving a mass parameter that depends on the wavevector direction $\hat{\bf k}$. Integrating out $k$ first in (\ref{Rate5}) leads to the interband Raman scattering rate
\begin{eqnarray}\label{Rate5b}
&& R'(\Omega) = \sqrt{2|\Omega|}\;\sum_{nn'} \int\frac{d^{2}\hat{\bf k}}{(2\pi)^{3}}\sum_{\sigma}\gamma_{\sigma n,-\sigma n'}^{\textrm{i,s}}(\hat{\bf k})\,\gamma_{-\sigma n',\sigma n}^{\textrm{s,i}}(\hat{\bf k}) \nonumber \\
&& \qquad\quad\times \left\vert\widetilde{m}_{\hat{\bf k}+}^{-1}\!-\!\widetilde{m}_{\hat{\bf k}-}^{-1}\right\vert^{-3/2}\, \theta\Bigl( |\Omega|\!-\!(E_{{\bf k}_{\textrm{F}}(\hat{\bf k})+}\!-\!E_{{\bf k}_{\textrm{F}}(\hat{\bf k})-}) \Bigr) \nonumber \\[0.1in]
&& \qquad\quad = \sqrt{|\Omega|} \, f(\hat{\bf e}_{\textrm{i}},\hat{\bf e}_{\textrm{s}},|\Omega|) \, \theta(|\Omega|-\Omega_0) \ ,
\end{eqnarray}
where the threshold frequencies are
\begin{eqnarray}
\Omega_0 &=& \min\left(E_{{\bf k}_{\textrm{F}}(\hat{\bf k})+}\!-\!E_{{\bf k}_{\textrm{F}}(\hat{\bf k})-}\right) \nonumber \\
\Omega_1 &=& \max\left(E_{{\bf k}_{\textrm{F}}(\hat{\bf k})+}\!-\!E_{{\bf k}_{\textrm{F}}(\hat{\bf k})-}\right) \nonumber
\end{eqnarray}
and the function $f$ becomes frequency-independent for $|\Omega|>\Omega_1$.

\begin{figure}
\subfigure[{}]{\includegraphics[height=1.70in]{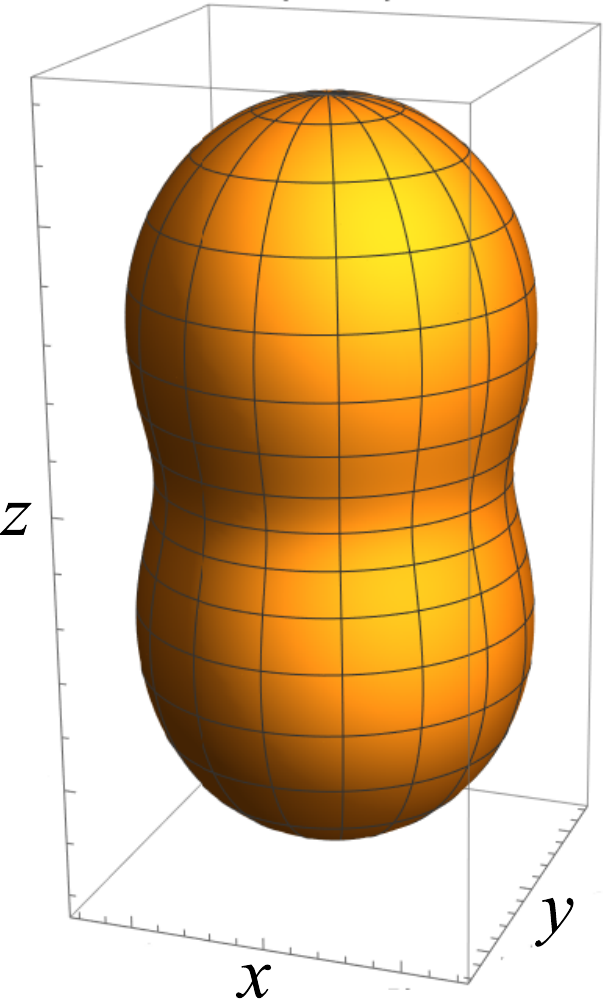}}
\subfigure[{}]{\includegraphics[height=1.60in]{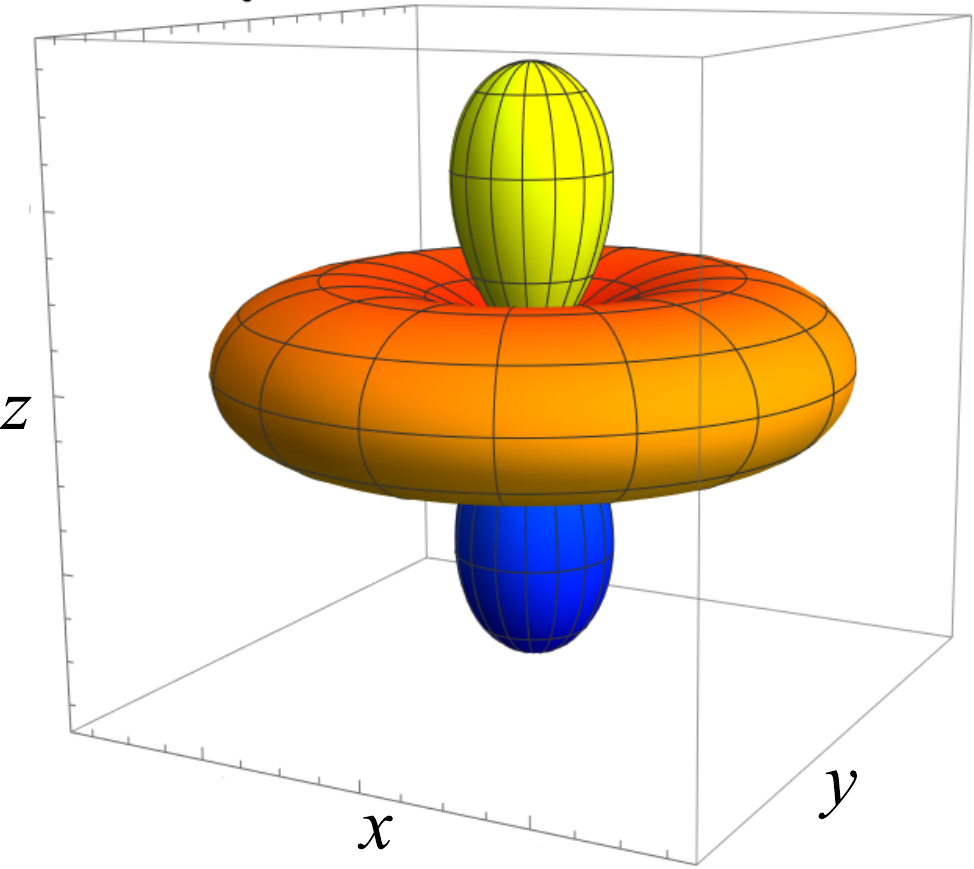}}
\subfigure[{}]{\includegraphics[width=1.70in]{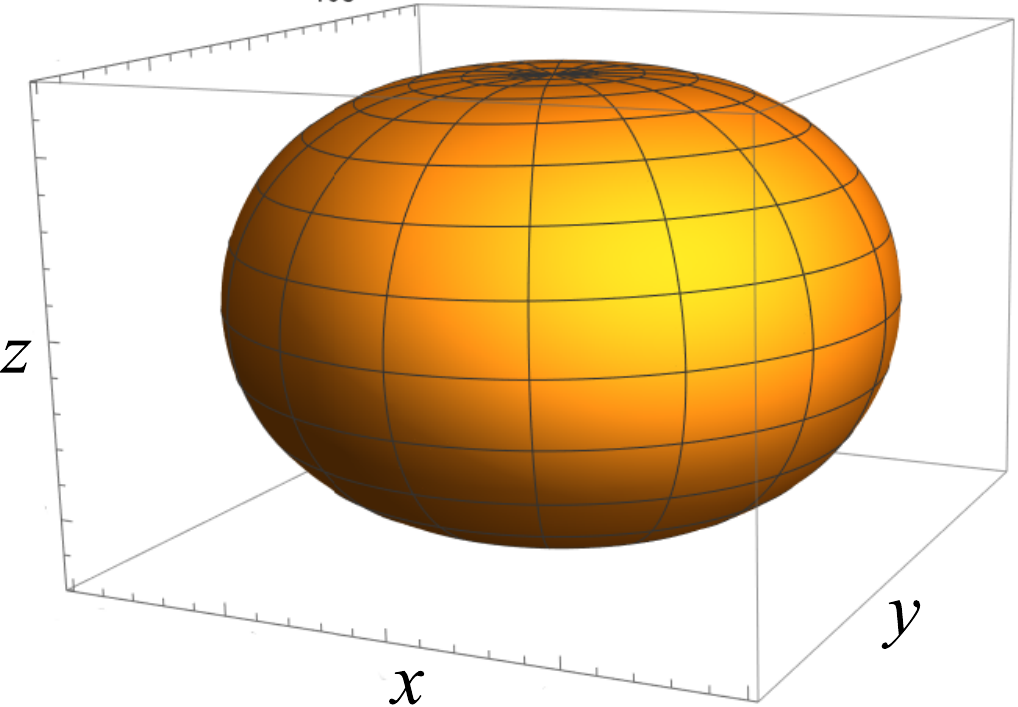}}
\subfigure[{}]{\includegraphics[width=1.60in]{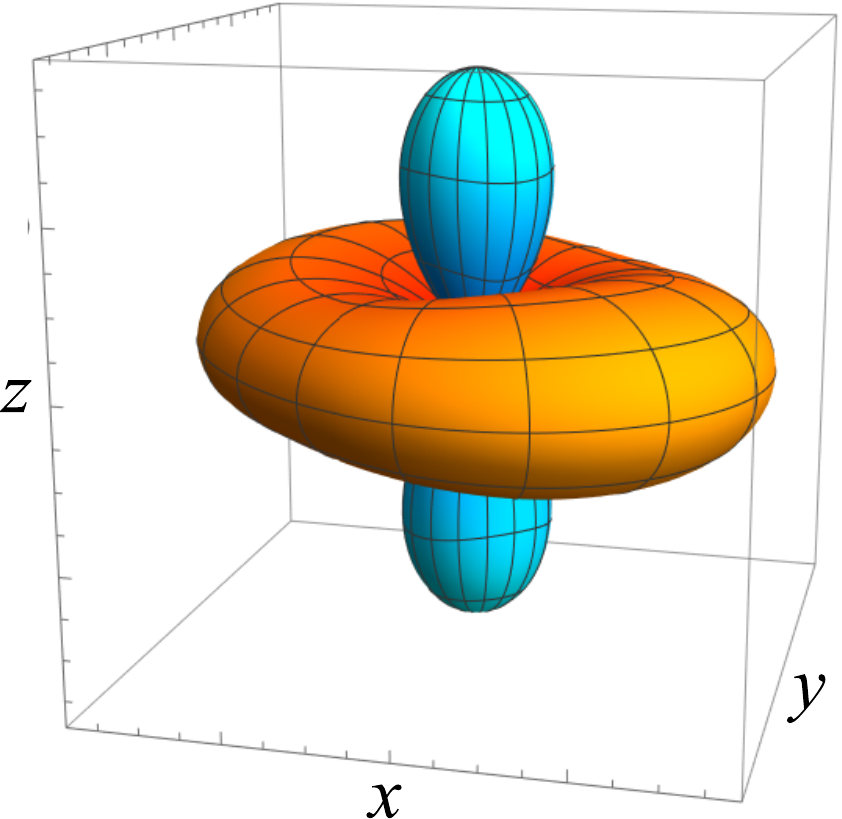}}
\subfigure[{}]{\includegraphics[width=1.65in]{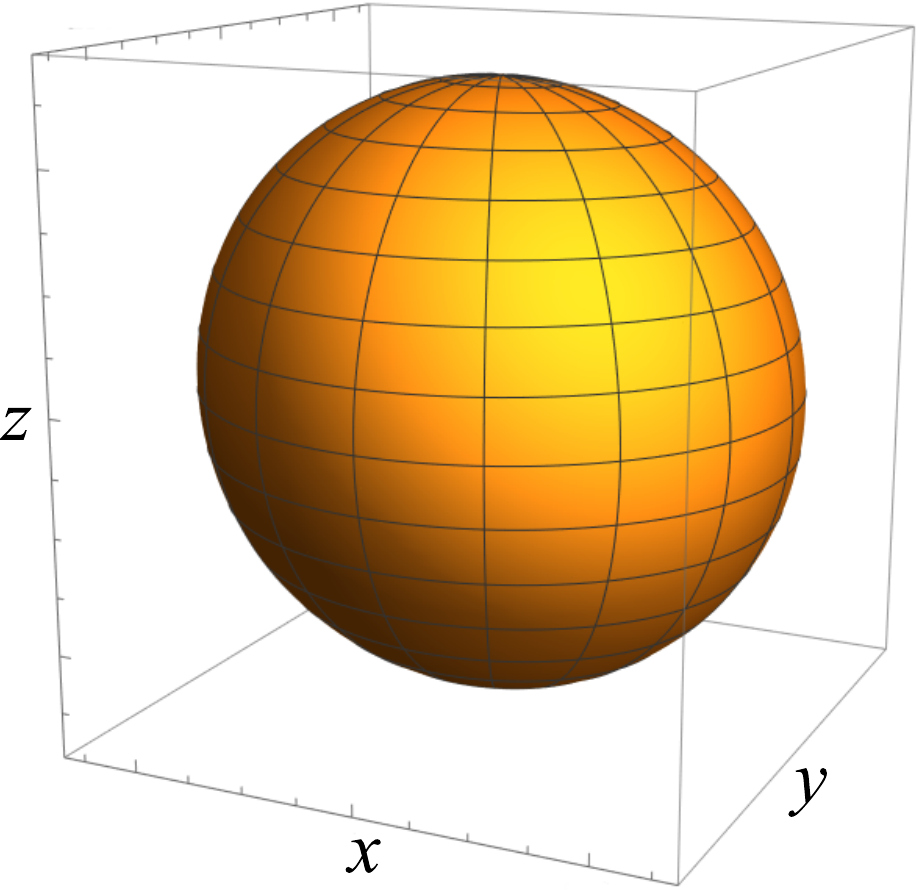}}
\subfigure[{}]{\includegraphics[width=1.65in]{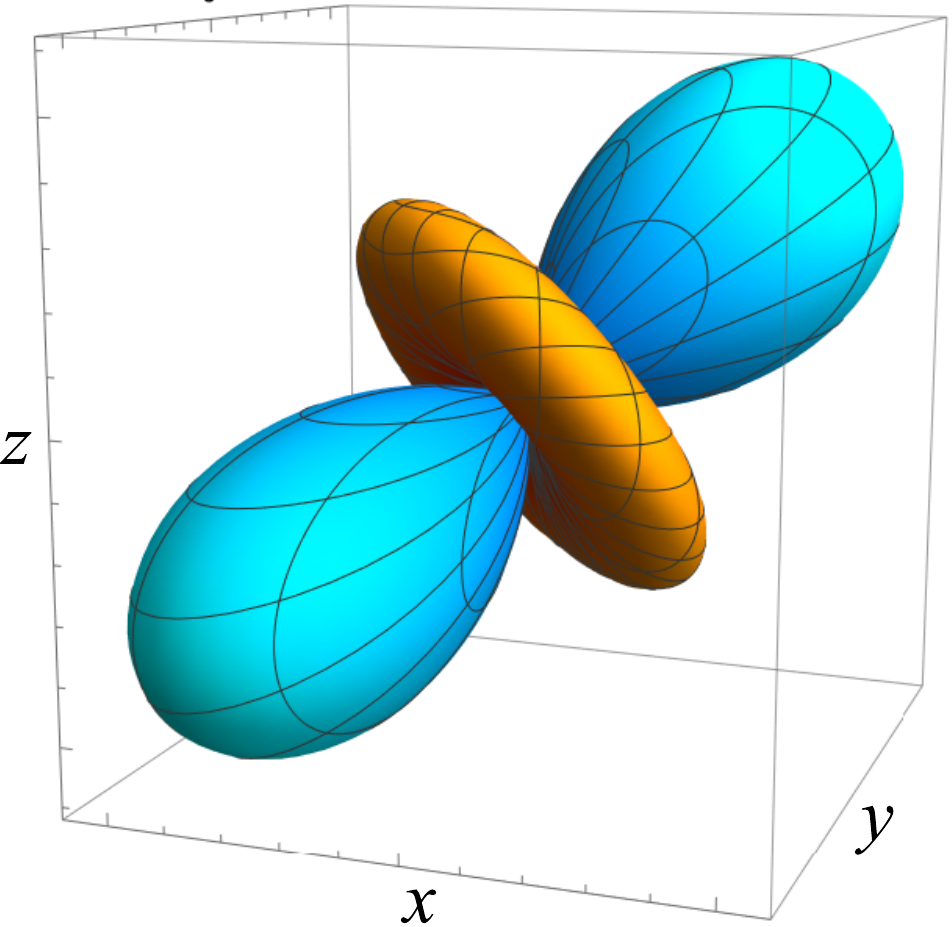}}
\caption{\label{LuttingerPol}The light polarization dependence $I_{\textrm{L}}(\hat{\bf e}_{\textrm{s}},\hat{\bf e}_{\textrm{i}})$ of the Raman scattering from nodal electrons with quadratic band touching. The incident light is assumed to have linear polarization in the direction (a,b) $\hat{\bf e}_{\textrm{i}}=(0,0,1)$, (c,d) $\hat{\bf e}_{\textrm{i}}=(1,1,0)/\sqrt{2}$ or (e,f) $\hat{\bf e}_{\textrm{i}}=(1,1,1)/\sqrt{3}$. The left column shows spherical plots of the Raman intensity in arbitrary units as a function of the direction of the scattered light polarization vector $\hat{\bf e}_{\textrm{s}}=(x,y,z)/\sqrt{x^2+y^2+z^2}$ for the given fixed incident light polarization $\hat{\bf e}_{\textrm{i}}$. The obtained surfaces reflect the cubic lattice symmetry and the directional bias set by the fixed $\hat{\bf e}_{\textrm{i}}$. The right column shows spherical plots of the difference $|I_{\textrm{L}}-r|$ between the Raman intensity $I_{\textrm{L}}$ and its best spherical fit with radius $r$ for each given incident polarization (color coding is red for $I_{\textrm{L}}>r$ and blue for $I_{\textrm{L}}<r$). This depicts the symmetries and spherical components: (a,b) $I_{\textrm{L}}\propto 1s+0.24d_{z^2}$, (c,d) $I_{\textrm{L}}\propto 1s-0.12d_{z^2}-0.01d_{xy}$, (e,f) $I_{\textrm{L}}\propto 1s-0.01(d_{xy}+d_{yz}+d_{zx})$, for the model parameters $m'/M=0.2$ and $M_{\textrm{c}}/M=0.6$ in this calculation. Note that the composition of non-zero spherical harmonics is qualitatively the same as in the case of Weyl electrons (Eq.\ref{PolarI1}) for each considered $\hat{\bf e}_{\textrm{i}}$, but the relative signed amplitudes of different harmonics are different within symmetry restrictions and depend here on the model parameters.}
\end{figure}

The dependence of the Raman scattering on the polarization of light is determined by the integral of Raman vertex functions over the electron's wavevector directions $\hat{\bf k}$ in (\ref{Rate5b}). Due to the inherent cubic symmetry of the model (\ref{HamiltonianLuttinger}), it is no longer easy to obtain an analytic formula, but a qualitative behavior can be deduced numerically. A good way to simplify the analysis is to neglect the cubic anisotropy of the spectrum, i.e. approximate $M_{\textrm{c}}\gg m'$ in (\ref{spectrumLuttinger}) while faithfully calculating the Raman vertex with the given finite $M_{\textrm{c}}$. This will let us focus on the polarization content without being distracted by the spreading of the Raman threshold into a finite range of frequencies. With this choice, the $\hat{\bf k}$ integral of (\ref{Rate5b}), which governs the interband transitions, reduces to
\begin{equation}
I_{\textrm{L}}(\hat{{\bf e}}_{\textrm{s}},\hat{{\bf e}}_{\textrm{i}})=\int d^{2}\hat{{\bf k}}\,\textrm{tr}\Bigl(P_{-}^{\phantom{\dagger}}\widetilde{\gamma}_{\hat{{\bf k}}}P_{+}^{\phantom{\dagger}}\widetilde{\gamma}_{\hat{{\bf k}}}+P_{+}^{\phantom{\dagger}}\widetilde{\gamma}_{\hat{{\bf k}}}P_{-}^{\phantom{\dagger}}\widetilde{\gamma}_{\hat{{\bf k}}}\Bigr) \ ,
\end{equation}
where an effective Raman vertex is given by
\begin{equation}
\gamma_{\sigma n, -\sigma n'}^{\textrm{i,s}} = \langle \sigma,n | \widetilde{\gamma}_{\bf k} | -\sigma, n'\rangle
\end{equation}
and
\begin{equation}
\widetilde{\gamma}_{\bf k} = -m\left( \frac{\lbrace\hat{{\bf e}}_{\textrm{i}}{\bf J},\hat{{\bf e}}_{\textrm{s}}{\bf J}\rbrace}{2m'} +\frac{\hat{e}_{\textrm{i}}^x \hat{e}_{\textrm{s}}^x J_x^2 + \hat{e}_{\textrm{i}}^y \hat{e}_{\textrm{s}}^y J_y^2 + \hat{e}_{\textrm{i}}^z \hat{e}_{\textrm{s}}^z J_z^2}{M_{c}} \right) \nonumber \ ,
\end{equation}
with projection operators
\begin{equation}
P_{\sigma}=\sum_{n}|\sigma, n\rangle\langle \sigma,n|
\end{equation}
to the lower ($\sigma=-1$) and upper ($\sigma=+1$) bands of the Luttinger node Hamiltonian (\ref{HamiltonianLuttinger}). The quantity $I_{\textrm{L}}(\hat{{\bf e}}_{\textrm{s}},\hat{{\bf e}}_{\textrm{i}})$ is plotted in Fig.\ref{LuttingerPol} for the incident light polarization vector $\hat{\bf e}_{\textrm{i}}$ pointing along the high-symmetry directions.

\begin{figure}
\includegraphics[width=3.2in]{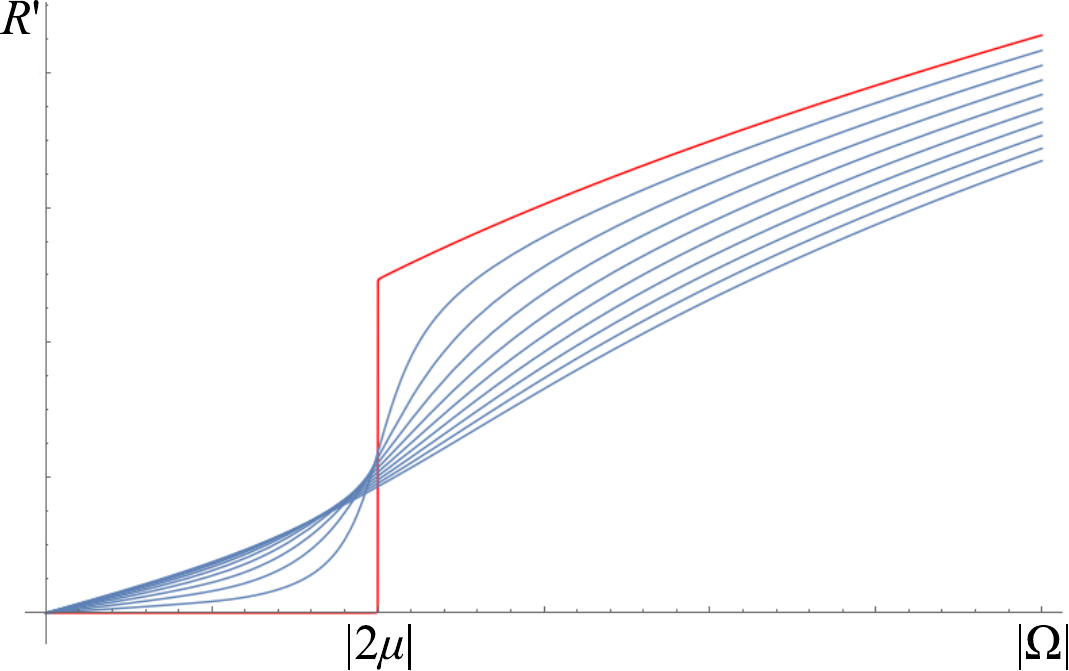}
\caption{\label{Linterband}Interband Raman scattering rate from the quasiparticles in a Luttinger semimetal, parametrized by the quasiparticle lifetime $\tau$. The Raman shift frequency $\Omega$ and the scattering rate $R'$ are shown in arbitrary units. The Luttinger spectrum was simplified to $E_{{\bf k}\sigma} = \sigma k^2/2\widetilde{m}-\mu$ for this example ($\sigma=\pm1$), with $2\widetilde{m}=1$ in the given units. The red curve corresponds to infinite lifetime -- scattering occurs only above a threshold frequency ($|2\mu|$ for particle-hole symmetric bands), and proceeds approximately as $R'\propto\sqrt{\Omega}$. The blue curves illustrate the evolution of Raman scattering as the quasiparticle ``decay rate'' $\Gamma\propto\tau^{-1}$ increases from $0.1$ in steps $0.1$ (expressed with the same units as $\Omega$).}
\end{figure}

The main benefit of this intuitive result is seeing that the frequency dependence of the interband Raman response correlates directly with the quasiparticle density of states. This is expected on physical grounds and observed also in the case of Weyl electrons. Now we consider the realistic consequences of having a finite quasiparticle lifetime. We must use the Green's functions (\ref{Green2}) with the spectrum (\ref{spectrumLuttinger}) to compute the expression analogous to (\ref{Rate3})
\begin{eqnarray}\label{Rate6}
R'(\Omega) &=& \frac{1}{\pi}\sum_{nn'}\int\frac{d^{3}k}{(2\pi)^{3}}\sum_{\sigma}\gamma_{\sigma n, -\sigma n'}^{\textrm{i,s}}({\bf k})\,\gamma_{-\sigma n', \sigma n}^{\textrm{s,i}}({\bf k}) \nonumber \\
&& \qquad \times \textrm{Re}\left\lbrace I_{\sigma}^{\phantom{x}}({\bf k},\Omega)\right\rbrace \ .
\end{eqnarray}
The function $\textrm{Re}\lbrace I_{\sigma} \rbrace$ is universally given by (\ref{ReIsigma}). Neglecting the detailed anisotropy of the cubic-symmetry spectrum and using the effective mass approximation allows us to again separate the $\hat{\bf k}$ and $k=|{\bf k}|$ integrations and extract the qualitative frequency dependence of the Raman response from the ensuing $\hat{\bf k}$-independent $I_\sigma$. The integral of $I_\sigma$ over $k$ needs to be calculated numerically but easily yields the frequency dependence $R'(\Omega)$ depicted in Fig.\ref{Linterband}.

The obtained results from the effective mass approximation take into account the virtual transitions of quasiparticles into high-energy bands, assuming that these band energies are much larger than the incoming and scattered photon energies $\omega_{\textrm{i}}$, $\omega_{\textrm{s}}$ respectively. The effective mass approximation breaks down for virtual transitions that occur within the nodal Luttinger spectrum. The contribution of such transitions to the Raman scattering rate must be computed using the complete expression (\ref{Gamma1}) for the vertex functions. To make the analysis simpler, we will ignore all high-energy bands and treat the effective theory (\ref{HamiltonianLuttinger}) as a microscopic model. The coupling of Luttinger quasiparticles to light is obtained by ``gauging'' the Hamiltonian, i.e. replacing the appearances of the wavevector components $k_i$ with the gauged momentum operators $p_i-\frac{e}{c}A_i$. The scattering vertex (\ref{Gamma1}) is constructed at the second-order perturbation theory with respect to the gauge field $A_i$, so that
\begin{eqnarray}\label{Gamma5}
&& \gamma_{\alpha\beta}^{i,s}=\widetilde{\rho}_{\alpha\beta}(\hat{{\bf e}}_{i},\hat{{\bf e}}_{s},{\bf q}_{i}-{\bf q}_{s})+\frac{1}{m}\sum_{\gamma} \\
&& \quad\times \left\lbrack \frac{\widetilde{p}_{\alpha\gamma}^{\phantom{x}}(\hat{{\bf e}}_{s}^{\phantom{\dagger}},\!-{\bf q}_{s}^{\phantom{\dagger}})\widetilde{p}_{\gamma\beta}^{\phantom{x}}(\hat{{\bf e}}_{i}^{\phantom{\dagger}},{\bf q}_{i}^{\phantom{\dagger}})}{E_{\beta}-E_{\gamma}+\omega_{i}}\!+\!\frac{\widetilde{p}_{\alpha\gamma}^{\phantom{x}}(\hat{{\bf e}}_{i}^{\phantom{\dagger}},{\bf q}_{i}^{\phantom{\dagger}})\widetilde{p}_{\gamma\beta}^{\phantom{x}}(\hat{{\bf e}}_{s}^{\phantom{\dagger}},\!-{\bf q}_{s}^{\phantom{\dagger}})}{E_{\beta}-E_{\gamma}-\omega_{s}}\right\rbrack \nonumber
\end{eqnarray}
contains the matrix elements in the Hamiltonian eigenstate $|\alpha,\beta\rangle$ basis
\begin{eqnarray}\label{Melements}
\widetilde{\rho}_{\alpha\beta}^{\phantom{x}}(\hat{{\bf e}}_{\textrm{i}},\hat{{\bf e}}_{\textrm{s}},{\bf q}) &\propto& \hat{e}_{\textrm{i}}^a\hat{e}_{\textrm{s}}^b\,\langle\alpha|e^{i{\bf q}{\bf r}}Q_{ab}|\beta\rangle \\
\widetilde{p}_{\alpha\beta}^{\phantom{x}}(\hat{{\bf e}},{\bf q}) &\propto& \hat{e}^a\,\langle\alpha|e^{i{\bf q}{\bf r}}P_{a}|\beta\rangle \ , \nonumber
\end{eqnarray}
where the operators $P_{a}$ and $Q_{ab}$ are
\begin{eqnarray}
P_{i} &=& \left(\frac{1}{M}+\frac{5}{4m'}\right)p_{i}-\sum_{j}\frac{p_{j}}{2m'}\lbrace J_{i},J_{j}\rbrace-\frac{J_{i}^{2}}{M_{c}}p_{i} \nonumber \\
Q_{ij}^{\phantom{x}} &=& \frac{1}{2}\left(\frac{1}{M}+\frac{5}{4m'}\right)\delta_{ij}^{\phantom{x}}-\frac{J_{i}J_{j}}{2m'}-\frac{J_{i}^{2}}{2M_{c}}\delta_{ij}^{\phantom{x}} \ .
\end{eqnarray}
Since the photon momenta ${\bf q}_{\textrm{i,s}}$ are negligible, we can compute the matrix elements of $P_i$ and $Q_{ij}$ with the Hamiltonian eigenstates $|\alpha\rangle$, $|\beta\rangle$ taken at the same wavevector ${\bf k}$. All matrix elements are then derived from
\begin{equation}
\mathcal{J}_{ij}^{(\alpha\beta)}({\bf k})=\langle\alpha|J_{i}^{\phantom{x}}J_{j}^{\phantom{x}}|\beta\rangle \ .
\end{equation}
The tensors $\mathcal{J}^{(\alpha\beta)}$ have a few notable properties: $\mathcal{J}^{(\alpha\beta)}=\left(\mathcal{J}^{(\beta\alpha)}\right)^{\dagger}$, $\mathcal{J}^{(\alpha\alpha)}=\left(\mathcal{J}^{(\beta\beta)}\right)^{\dagger}$ if $\alpha$ and $\beta$ are two orthogonal states from the same band. Importantly, $\mathcal{J}_{ij}^{(\alpha\beta)}$ is independent of $|{\bf k}|$. Its dependence on $\hat{\bf k}$ reflects the cubic symmetry and obeys $\mathcal{J}_{ij}^{(\alpha\beta)}(\hat{{\bf k}})=\mathcal{J}_{ij}^{(\alpha\beta)}(-\hat{{\bf k}})$.

Considering these properties, it is evident that the Raman vertex functions (\ref{Gamma5}) carry momentum dependence in the form of
\begin{equation}
\gamma_{\alpha\beta}^{\textrm{i,s}} = a(\hat{\bf k}) + \frac{k^2}{m}\,b(\hat{\bf k})
\end{equation}
in the present model. The first term is contained in the effective mass approximation (\ref{Gamma4}) when the proportionality constants of (\ref{Melements}) are recovered, but the second term is a $k$-dependent correction. The correction is small in the limit $\Omega\ll \omega_{\textrm{i}},\omega_{\textrm{s}}$ because $\Omega \propto k^2$ in interband transitions by energy conservation and $b(\hat{\bf k}) \propto \omega_{\textrm{i,s}}^{-1}$. Therefore, we can rely on the qualitative results obtained earlier at least in the $\Omega\ll \omega_{\textrm{i}},\omega_{\textrm{s}}$ limit. Obviously, the correction due to $b(\hat{\bf k})$ can alter the frequency response of the interband scattering rate $R'$ if it becomes large enough. However, even then, only additional even powers of $k$ are introduced in the integrals of (\ref{Rate5}) and (\ref{Rate5b}), giving rise to a frequency dependence that can be expanded as
\begin{equation}
R'(\Omega) \sim \sum_n r_n |\Omega|^{\frac{1}{2}+n}
\end{equation}
over positive integers $n$. None of the terms in this expansion can reproduce the $R'\sim\Omega^2$ response of Weyl electrons, so a careful analysis of the experimental data can still tell a difference between the Weyl and Luttinger quasiparticles.

At this point, it remains only to include the intraband scattering in the total Raman response. Intraband processes contribute only if the quasiparticles have a finite lifetime. We will restrict the analysis to the effective mass approximation and neglect the detailed cubic anisotropy of the spectrum, particle-hole asymmetry of the bands, etc. The simplified spectrum $E_{k\sigma} = \sigma k^2/2\widetilde{m}-\mu$ is good enough for reaching qualitative conclusions and extracting the scale of the correction. Following the steps we took in the case of Weyl quasiparticles, we find that the intraband scattering ``rate'' has the form analogous to (\ref{Rate4b})
\begin{eqnarray}
R''(\Omega) &\approx& \frac{1}{4\pi^{4}}\,\frac{2\Gamma}{|\Omega|^{2}+4\Gamma^{2}} (2\widetilde{m}\Gamma)^{\frac{3}{2}} I''\left(\frac{|\Omega|}{\Gamma}, \frac{\mu}{\Gamma}\right) \\
&& \times \sum_{nn'} \int d^{2}\hat{\bf k}\,\sum_{\sigma}\gamma_{\sigma n,\sigma n'}^{\textrm{i,s}}(\hat{\bf k})\,\gamma_{\sigma n',\sigma n}^{s,i}(\hat{\bf k})  \ . \nonumber
\end{eqnarray}
The dimensionless function $I''(\phi,\nu)$ is given by (\ref{Ipp}) but with a dimensionless energy
\begin{equation}
\xi_{k\sigma} = \frac{E_{k\sigma}}{\Gamma} = \sigma\frac{k^2}{2\Gamma\widetilde{m}} - \frac{\mu}{\Gamma} = \sigma\kappa^2 - \nu
\end{equation}
appropriate for the Luttinger spectrum. Since we can expand $I''(\phi,\nu)\propto(1+\xi^2)^{-1}\phi + \mathcal{O}(\phi^3)$ for small frequencies $|\Omega|=\Gamma\phi$, the overall frequency dependence of the intraband Raman scattering ``rate'' is again
\begin{equation}
R''(\Omega) \sim \frac{\tau|\Omega|}{\tau^2\Omega^2+1} \ ,
\end{equation}
where $\tau\propto\Gamma^{-1}$.

\section{Dirac spectrum and a flat band}\label{Flat}

Flat bands are another feature of the electronic spectrum that often appears in conjunction with non-trivial topology, specifically the Hall effect. It is possible, in principle, to detect the presence of flat bands with Raman scattering, and here we explore the characteristic universal features of Raman response that could indicate flat bands. Specifically, we consider a two-band scenario in which photons can induce transitions between a quasi two-dimensional Dirac band and a flat band over a finite region of the first Brillouin zone. A minimal Hamiltonian has the form
\begin{equation}
H_{{\bf k}}=\mathcal{P}_{-}\frac{({\bf p}-\sigma^{a}{\bf A}^{a})^{2}}{2m}\mathcal{P}_{-}+\mathcal{P}_{+}\epsilon\mathcal{P}_{+}-\mu' \ .
\end{equation}
The operators $\mathcal{P}_{\pm} = (1\pm s^z)/2$ are projections onto the Dirac and flat-band subspaces. The eigenvalue of the Pauli operator $s^z$ is the band index which identifies the $s^{z}=-1$ Dirac branch and the $s^{z}=+1$ flat band with energy $\epsilon$. The spin degrees of freedom are handled with the Pauli operators $\sigma^a$, and the gauge field ${\bf A}^{a}$ given by (\ref{GaugeField}), together with (\ref{ChemPot}), forges a massless Dirac spectrum with Fermi level placed at energy $\mu$ away from the node. We are, in fact, modeling the Dirac spectrum as a pair of Weyl nodes that coincide at the same wavevector but have opposite chiralities. The resulting two-fold degeneracy of the Dirac spectrum is not modeled here (the two degenerate branches make the same contribution to the Raman signal), and we assume for the calculations that the Dirac point is not gapped out. The Green's function we need still obtains from the Hamiltonian as (\ref{Green1}).

\begin{figure}
\includegraphics[height=1.5in]{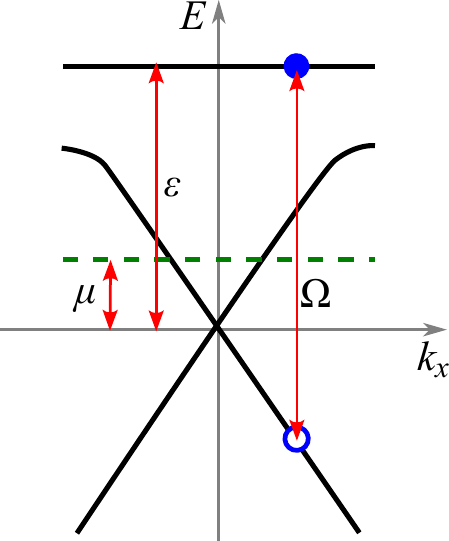}
\caption{\label{Wayl-flat}Illustration of a non-resonant Raman process involving a Dirac and a flat band.}
\end{figure}

In computing the Raman vertex (\ref{Gamma1}), we again need to go beyond the effective mass approximation in order to capture the Raman scattering cross-section from a Dirac and a flat band. We will consider the case of the initial and final states $|\alpha\rangle,|\beta\rangle$ being in different bands, one in the Dirac band and the other in the flat band. The intermediate state $|\gamma\rangle$  also needs to be in the Dirac or flat band, at least in order to produce the dominant contribution (given $E_{\alpha/\beta}-E_{\gamma}$ in the denominators of the Raman vertex and a nearly resonant photon frequency for the Dirac-flat interband transitions). Labeling the Dirac and flat band states as $|\delta\rangle\equiv|\sigma{\bf k}\rangle$ and $|\phi\rangle\equiv|\widetilde{\sigma}{\bf k}\rangle$ respectively, the interband transitions are characterized by:
\begin{eqnarray}
\langle\delta_{\sigma{\bf k}}|e^{i{\bf q}{\bf r}}|\delta_{\sigma'{\bf k}'}\rangle &=& \delta_{\sigma\sigma'}\delta_{{\bf k},{\bf k}'+{\bf q}} \\
\langle\phi_{\widetilde{\sigma}{\bf k}}|e^{i{\bf q}{\bf r}}|\phi_{\widetilde{\sigma}'{\bf k}'}\rangle &=& \delta_{\widetilde{\sigma}\widetilde{\sigma}'}\delta_{{\bf k},{\bf k}'+{\bf q}} \nonumber \\\langle\phi_{\widetilde{\sigma}{\bf k}}|e^{i{\bf q}{\bf r}}|\delta_{{\bf \sigma'k}'}\rangle &\approx& 0 \nonumber \ .
\end{eqnarray}
Neglecting momentum transfers ${\bf q}\to0$, we also find
\begin{equation}
\hat{{\bf e}}\,\langle\phi_{\widetilde{\sigma}{\bf k}}|e^{i{\bf q}{\bf r}}\boldsymbol{\pi}|\phi_{\widetilde{\sigma}'{\bf k}'}\rangle\approx0
\end{equation}
\vspace{-0.3in}
\begin{eqnarray}
\hat{{\bf e}}\,\langle\delta_{\sigma{\bf k}}|e^{i{\bf q}{\bf r}}\boldsymbol{\pi}|\delta_{\sigma'{\bf k}'}\rangle
  &\!=\!& \hat{{\bf e}}\,\delta_{{\bf k},{\bf k}'}\bigl({\bf k}\delta_{\sigma\sigma'}\!+\!mv\langle\sigma_{{\bf k}}|\boldsymbol{\sigma}|\sigma'_{{\bf k}}\rangle\bigr) \nonumber \\
\hat{{\bf e}}\,\langle\phi_{\widetilde{\sigma}{\bf k}}|e^{i{\bf q}{\bf r}}\boldsymbol{\pi}|\delta_{{\bf \sigma'k}'}\rangle
  &\!=\!& \hat{{\bf e}}\,\delta_{{\bf k},{\bf k}'}\bigl(\kappa_{p}{\bf k}\delta_{\widetilde{\sigma}\sigma'}\!+\!mv\kappa_{\sigma}\langle\widetilde{\sigma}_{{\bf k}}|\boldsymbol{\sigma}|\sigma'_{{\bf k}}\rangle\bigr) \nonumber \\
\hat{{\bf e}}\,\langle\delta_{{\bf \sigma k}}|e^{i{\bf q}{\bf r}}\boldsymbol{\pi}|\phi_{\widetilde{\sigma}'{\bf k}'}\rangle
  &\!=\!& \hat{{\bf e}}\,\delta_{{\bf k},{\bf k}'}\bigl(\kappa_{p}{\bf k}\delta_{\sigma\widetilde{\sigma}'}\!+\!mv\kappa_{\sigma}\langle\sigma_{{\bf k}}|\boldsymbol{\sigma}|\widetilde{\sigma}'_{{\bf k}}\rangle\bigr) \nonumber
\end{eqnarray}
where $\boldsymbol{\pi}$ is the effective gauged momentum operator that couples to the vector potential of the U(1) gauge field in the first order perturbation theory. The modeling of this operator proceeds after extracting the overall $1/m$ factor in (\ref{Gamma1}) given by the non-relativistic mass $m\equiv m_{\textrm{Dirac}}$ in the background of the Dirac spectrum. Each band-diagonal term of $\boldsymbol{\pi}$ introduces a $1/m_{n}$ factor in (\ref{Gamma1}), where $m_{n}$ is the effective mass in the band $n$; since the flat band has an extremely large effective mass, its diagonal term in $\boldsymbol{\pi}$ is by a factor of $m_{\textrm{Dirac}}/m_{\textrm{flat}}\to0$ smaller than the included diagonal Dirac band's term. Furthermore, the U(1) vector potential is generally not diagonal in the band eigenbasis, and thus enables inter-band optical transitions. In the low-energy sector, the flat band introduces no bias for the momentum and spin dependence of the inter-band matrix elements, so we are justified in using the universal form for the inter-band matrix elements that stems from the Dirac spectrum. We only need to introduce two phenomenological constants $\kappa_{p}$ and $\kappa_{\sigma}$ whose values can be determined by solving the ab-initio band structure. The flat band must be spin-degenerate in a time-reversal-invariant state, or Zeeman-split in ferromagnetic phases (which we will not consider here). Hence, we label the spin content of the flat band eigenstates with $\widetilde{\sigma}$ and treat the spin and momentum in the flat band as independent and uncorrelated quantum numbers. Effectively, the spin-orbit gauge field vanishes in the flat band by the lack of spin-momentum correlation (if we kept it in, the SU(2) gauge flux $m_{\textrm{flat}}v_{\textrm{flat}}$ and the chemical potential shift $m_{\textrm{flat}}^{\phantom{x}}v_{\textrm{flat}}^{2}$ would diverge for any finite $v_{\textrm{flat}}$).

In the subsequent analysis, we are free to pick the same momentum-dependent spin basis $\widetilde{\sigma}$ for the flat-band and the Dirac band. The Dirac electron states at $\hat{{\bf k}}=\hat{{\bf x}}\sin\theta\cos\phi+\hat{{\bf y}}\sin\theta\sin\phi+\hat{{\bf z}}\cos\theta$ are given by (\ref{SpinCoherentStates}). In the case of a 2D Dirac spectrum, $\theta=\pi/2$, we have
\begin{eqnarray}
\langle\sigma_{\bf k}|\boldsymbol{\sigma}|\sigma_{\bf k}\rangle &=& \sigma\hat{\bf k} \\
\langle\sigma_{\bf k}|\boldsymbol{\sigma}|-\sigma_{\bf k}\rangle &=& -e^{-i\sigma\phi}\lbrack i\sigma(\hat{\bf z}\times\hat{\bf k})+\hat{\bf z}\rbrack \ , \nonumber
\end{eqnarray}
The Raman vertex (\ref{Gamma1}) for infinite quasiparticle lifetime, where the conservation of energy implies $\omega_{s}-\omega_{i}=\epsilon-\sigma vk$, is
\begin{widetext}
\begin{eqnarray}
&& \gamma_{\delta\phi}^{\textrm{i,s}} \approx
\frac{\delta_{{\bf k}{\bf k}'}}{m}\delta_{\sigma\sigma'}\biggl\lbrack \frac{(\hat{{\bf e}}_{\textrm{i}}\hat{{\bf k}})(\hat{{\bf e}}_{\textrm{s}}\hat{{\bf k}})( k+\sigma mv)(\kappa_{p} k+\kappa_{\sigma}\sigma mv)}{\omega_{\textrm{s}}} +\frac{\kappa_{\sigma}(mv)^{2}(\hat{{\bf e}}_{\textrm{i}}\langle-\!\sigma_{{\bf k}}|\boldsymbol{\sigma}|\sigma_{{\bf k}}\rangle)(\hat{{\bf e}}_{\textrm{s}}\langle\sigma_{{\bf k}}|\boldsymbol{\sigma}|-\!\sigma_{{\bf k}}\rangle)}{\omega_{\textrm{s}}}-(\cdots)_{\textrm{i}\leftrightarrow \textrm{s}}\biggr\rbrack \\
&& \qquad~ +\frac{\delta_{{\bf k}{\bf k}'}}{m}(1\!-\!\delta_{\sigma\sigma'})\biggl\lbrack \frac{\kappa_{\sigma}mv(\hat{{\bf e}}_{\textrm{i}}\langle\sigma_{{\bf k}}|\boldsymbol{\sigma}|-\!\sigma_{{\bf k}}\rangle)(\hat{{\bf e}}_{\textrm{s}}\hat{{\bf k}})( k+\sigma mv)}{\omega_{\textrm{s}}} + \frac{mv(\hat{{\bf e}}_{\textrm{i}}\hat{{\bf k}})(\hat{{\bf e}}_{\textrm{s}}\langle\sigma_{{\bf k}}|\boldsymbol{\sigma}|-\!\sigma_{{\bf k}}\rangle)(\kappa_{p} k-\kappa_{\sigma}\sigma mv)}{\omega_{\textrm{s}}}-(\cdots)_{\textrm{i}\leftrightarrow \textrm{s}}\biggr\rbrack \nonumber
\end{eqnarray}
\end{widetext}
We also have $\gamma_{\phi\delta}^{\textrm{s,i}} = (\gamma_{\delta\phi}^{\textrm{i,s}})^*$. The scattering rate in the Raman processes contributed by interband transitions derives from the product of vertex functions
\begin{equation}
X_{\sigma\sigma'}^{\phantom{x}}({\bf k})\equiv\gamma_{\delta\phi}^{\textrm{i,s}}\gamma_{\phi\delta}^{\textrm{s,i}}
\end{equation}
and
\begin{equation}
\delta\chi(0,\Omega) = -i\sum_{{\bf k}\omega}\sum_{\sigma\sigma'}X_{\sigma\sigma'}({\bf k})\,G_{\sigma}({\bf k},\omega)\,G_{\sigma'}({\bf k},\omega+\Omega) \ .
\end{equation}
After some algebra, we finally obtain
\begin{widetext}
\begin{eqnarray}
&& \delta R = -\frac{1}{\pi}\textrm{Im}\left\lbrace \delta\chi(,\Omega)\right\rbrace \xrightarrow{\epsilon>\mu} \frac{\theta(\mu-\epsilon+\Omega)}{(2\pi)^{2} v} \int\limits_{0}^{2\pi}d\phi\left\lbrack \sum_{\sigma'}X_{\textrm{sign}(\epsilon-\Omega),\sigma'}\left(\frac{|\epsilon-\Omega|}{ v},\hat{{\bf k}}(\phi)\right)\right\rbrack \frac{|\epsilon-\Omega|}{ v} \\
\nonumber
&& \quad = \frac{|\epsilon-\Omega|\,\theta(\mu-\epsilon+\Omega)}{(2\pi)^{2}v^2} \int\limits_{0}^{2\pi}d\phi \nonumber \\
&& \quad\times\Biggl\lbrack \frac{1}{m^{2}}\left\vert \frac{1}{v^{2}}\frac{(\hat{{\bf e}}_{\textrm{i}}\hat{{\bf k}})(\hat{{\bf e}}_{\textrm{s}}\hat{{\bf k}})(|\epsilon\!-\!\Omega|\!+\!\sigma mv^{2})(\kappa_{p}|\epsilon\!-\!\Omega|\!+\!\kappa_{\sigma}\sigma mv^{2})}{\omega_{\textrm{s}}}+\frac{\kappa_{\sigma}(mv)^{2}(\hat{{\bf e}}_{\textrm{i}}\langle-\sigma_{{\bf k}}|\boldsymbol{\sigma}|\sigma_{{\bf k}}\rangle)(\hat{{\bf e}}_{\textrm{s}}\langle\sigma_{{\bf k}}|\boldsymbol{\sigma}|-\sigma_{{\bf k}}\rangle)}{\omega_{\textrm{s}}}-(\cdots)_{\textrm{i}\leftrightarrow \textrm{s}}\right\vert^{2} \nonumber \\
&& \qquad\quad +\left\vert \frac{\kappa_{\sigma}(\hat{{\bf e}}_{\textrm{i}}\langle\sigma_{{\bf k}}|\boldsymbol{\sigma}|-\sigma_{{\bf k}}\rangle)(\hat{{\bf e}}_{\textrm{s}}\hat{{\bf k}})(|\epsilon\!-\!\Omega|\!+\!\sigma mv^{2})}{\omega_{\textrm{s}}}+\frac{(\hat{{\bf e}}_{\textrm{i}}\hat{{\bf k}})(\hat{{\bf e}}_{\textrm{s}}\langle\sigma_{{\bf k}}|\boldsymbol{\sigma}|-\sigma_{{\bf k}}\rangle)(\kappa_{p}|\epsilon\!-\!\Omega|\!-\!\kappa_{\sigma}\sigma mv^{2})}{\omega_{\textrm{s}}}-(\cdots)_{\textrm{i}\leftrightarrow \textrm{s}}\right\vert^{2} \Biggr\rbrack \nonumber \ .
\end{eqnarray}
\end{widetext}
The angle integral contains some residual shift-frequency dependence through factors of the form
\begin{equation}
a|\epsilon-\Omega|+b\sigma mv^{2}\xrightarrow{mv^{2}\gg\epsilon,\Omega} bcmv^{2} \ .
\end{equation}
However, if the Dirac spectrum is very linear in the span of energies comparable to $\epsilon$, then the mass parameter $m$ is effectively extremely large and so $mv^{2}\gg\epsilon\sim\Omega$. This approximation impacts only the polarization dependence of the Raman scattering at leading order. The frequency dependence is given by
\begin{equation}
\delta R(\Omega)\approx \frac{|\epsilon-\Omega|\,\theta(\mu-\epsilon+\Omega)}{\pi(2\pi)^{2}(\hbar v)^{2}}\int\limits_{0}^{2\pi}d\phi\,X\!\left(\hat{{\bf k}}(\phi)\right) \ .
\end{equation}
Depending on the chemical potential $\mu$, this function of frequency can exhibit an ``hourglass'' feature as the shift frequency $\Omega$ passes through the energy separation $\epsilon$ between the flat band and the Dirac node. This characteristic indication of the coexisting flat and Dirac bands in the electron spectrum enables the experimental extraction of the flat band's energy $\epsilon$. If the Dirac node is gapped out due to the absence of a protective symmetry, then a further suppression of $\delta R(\Omega)$ in a finite frequency range near $\Omega\sim \epsilon \pm \Delta/2$ is expected.

\section{Discussion and conclusions}\label{Concl}

We presented the derivation of the cross-section for the Raman scattering on itinerant electrons with Weyl and quadratic band touching spectra. The relevant physical process is non-resonant inelastic photon scattering which takes an electron from an initial to a final state within the nodal spectrum, without a significant momentum transfer. In ideal circumstances, nodal Raman scattering is activated above a threshold frequency, which is proportional to the energy difference between the Fermi level and the node. All nodes in the first Brillouin zone contribute, but those that live at different energies can be, in principle, resolved at different frequencies. The frequency dependence of the Raman signal above the threshold, $R(\Omega)\propto \Omega^a$, is qualitatively related to the nodal electron density of states $\rho(E)\propto E^a$. In the case of Weyl electrons, the relativistic and massless spectrum gives rise to a universal frequency, $R(\Omega)\propto \Omega^2$, and polarization dependence of the Raman cross-section, which can be captured by an analytic expression when electrons have infinite lifetime. Likewise, quadratic band touching produces $R(\Omega)\propto \Omega^{1/2}$ in three-dimensional systems. Analogous connection to the density of states can be established for Dirac quasi-2D dispersions. A finite lifetime caused by interactions, disorder and thermal fluctuations, is easy to model with a single lifetime parameter $\tau$, and generally blurs the onset of the Raman signal across the threshold frequency into a continuous frequency dependence. Nevertheless, the expressions we derived can be easily numerically integrated and provide a degree of universality which allows fitting to experimental measurements. Overall, the measured frequency dependence of the Raman signal can be used to estimate the Fermi energy, Fermi velocity (determined by the strength of the spin-orbit coupling in the case of Weyl electrons), as well as the lifetime $\tau$.

The picture becomes more complex when the Weyl nodes have a tilted spectrum, as generally permitted by lattice symmetries in solids. We have derived the Raman scattering cross-section for the case of type-I nodes, and shown that the threshold frequency splits into two characteristic frequencies. If electrons have infinite lifetime, then the lower frequency is still a threshold for the onset of the Raman signal, while the Raman signal retains its universal ``untilted'' form above the higher frequency. In between, the Raman scattering cross-section evolves smoothly in a non-universal manner. Since the difference between the two split frequencies depends on the amount of tilt in the Weyl spectrum, one can in principle determine the tilt by resolving both frequencies. Of course, this is made more difficult in realistic circumstances with a finite lifetime, but specific formulas for the cross section are still easy to numerically integrate and fit to data. It should be noted that lattice symmetries generally constrain all Weyl nodes with the same node energy in the same manner, so a single well-defined tilt parameter can be extracted from the frequency dependence of the Raman scattering rate $R(\Omega)$.

Another new result in this study is the polarization dependence of the Raman signal. The $R(\Omega)\propto \Omega^2$ frequency dependence is universal above the upper threshold frequency in the case of Weyl electrons, and unburdened by detailed lattice symmetries. This is the part of the non-resonant Raman scattering forged exclusively by the Weyl spectrum and calculated without the effective mass approximation. The non-universal contributions involving higher-energy bands superimpose different powers of frequency in $R(\Omega)$, which, therefore, are distinguishable from the universal part in a fitting procedure. The provided analytic formula can then help identify and verify the observation of Weyl electrons in Raman experiments. We also derived analogous expression for the polarization dependence of the Raman scattering on electrons with quadratic band touching. In this case, the polarization dependence exhibits evidence of the symmetries that protect the quadratic band touching in the first place.

Finally, we also considered Raman scattering from quasi-two-dimensional electronic spectra that contain Dirac nodes and flat bands. Spectra such as these are often found for itinerant electrons in ``frustrated'' lattices with complex unit-cells, most famously the kagome lattice. Some candidate materials that can be studied with Raman scattering, such as V$_{1/3}$NbS$_2$, also possibly feature spectra of this kind. We have shown that non-resonant photons can drive virtual transitions between Dirac branches and a flat band, leading to a characteristic detectable frequency dependence of the Raman signal that traces the ``hourglass'' shape of the Dirac spectrum.

\bigskip
\section{Acknowledgments}\label{secAck}

I am very grateful to Natalia Drichko for earlier collaboration and ongoing insightful discussions. This research was supported at the Institute for Quantum Matter, an Energy Frontier Research Center funded by the U.S. Department of Energy, Office of Science, Basic Energy Sciences under Award No. DE-SC0019331.

\appendix
\bigskip

\section{A model for high-energy corrections in the Raman scattering on Weyl electrons}\label{appWeyl}

The Weyl spectrum deviates from a perfect linear form at high energies. This introduces certain non-universal corrections in the frequency dependence of the Raman signal. Here we consider a simple natural model that produces such corrections, and argue that the universal nodal contributions to the Raman signal are still qualitatively distinct and recognizable despite contamination from high energies.

We will scrutinize the electron Hamiltonian
\begin{equation}
H_{{\bf k}} = \frac{v}{a_{0}}\sigma^{a}\sin(a_{0}k^{a})-\mu
\end{equation}
which produces a spherically-symmetric untilted Weyl spectrum at low energies and a cubic distortion from the linear form at higher energies controlled by the lattice constant $a_0$,
\begin{equation}
\epsilon_{{\bf k}}=\frac{v}{a_{0}}\sqrt{\sum_{a}\sin^{2}(a_{0}k^{a})}\xrightarrow{{\bf k}\to0}v|{\bf k}|
\end{equation}
The summation over the repeated index $a\in\lbrace x,y,z \rbrace$ is assumed. This model has eight Weyl nodes in the first Brillouin zone at
\begin{equation}
{\bf Q} = \left(\frac{1-s_{x}}{2},\frac{1-s_{y}}{2},\frac{1-s_{z}}{2}\right)\frac{\pi}{a_{0}}
\end{equation}
labeled by $s_x, s_y, s_z = \pm1$, with node chiralities $s_{x}s_{y}s_{z}=\pm1$. The non-linearity of the spectrum enables the use of the effective mass approximation (\ref{Gamma1b}) for the Raman vertex:
\begin{equation}
\gamma_{{\bf k}}^{\phantom{x}}\approx m\hat{e}_{i}^{a}\hat{e}_{s}^{b}\frac{\partial^{2}H_{{\bf k}}}{\partial k^{a}\partial k^{b}} = -mva_{0}\sum_{c}\hat{e}_{i}^{c}\hat{e}_{s}^{c}\sigma^{c}\sin(a_{0}k^{c}) \ .
\end{equation}
The Raman susceptibility (\ref{Chi1}) is then:
\begin{widetext}
\begin{eqnarray}
&& \chi({\bf q},\Omega) = -i\left(mva_{0}\right)^{2}\sum_{cd}\hat{e}_{i}^{c}\hat{e}_{s}^{c}\hat{e}_{i}^{d}\hat{e}_{s}^{d}\int\limits_{\textrm{1BZ}}\!\!\frac{d^{3}k}{(2\pi)^{3}}\sin(a_{0}k^{c})\sin(a_{0}k^{d}+a_{0}q^{d}) \\
&& \qquad\times\int\frac{d\omega}{2\pi}\,\textrm{tr}\left\lbrace \frac{1}{\omega-\frac{v}{a_{0}}\sigma^{a}\sin(a_{0}k^{a})+\mu+i0^{+}\textrm{sign}(\omega)}\sigma^{c}\frac{1}{\omega+\Omega-\frac{v}{a_{0}}\sigma^{b}\sin(a_{0}k^{b}+a_{0}q^{b})+\mu+i0^{+}\textrm{sign}(\omega+\Omega)}\sigma^{d}\right\rbrace \nonumber
\end{eqnarray}
\end{widetext}
The subsequent calculations are tedious but proceed in the same fashion as before: one first evaluates the trace, then integrates the frequency $\omega$. Some approximations can be made because momentum transfers ${\bf q}$ from photons to electrons are small, and we are focusing on the case of infinite electron lifetime,
\begin{widetext}
\begin{eqnarray}
&& \chi({\bf q}\to0,\Omega) \approx \frac{1}{\Omega} \left(mva_{0}\right)^{2}\sum_{cd}\hat{e}_{i}^{c}\hat{e}_{s}^{c}\hat{e}_{i}^{d}\hat{e}_{s}^{d}\int\limits_{\textrm{1BZ}}\!\!\frac{d^{3}k}{(2\pi)^{3}}\,\frac{1}{2\epsilon_{{\bf k}}}\,\sin(a_{0}k^{c})\sin(a_{0}k^{d}) \\
&& \quad\times \Biggl\lbrace\frac{(X_{{\bf k}}^{cd}\!+\!2\epsilon_{{\bf k}}^{\phantom{x}}\Omega\delta^{cd})\,\theta(\mu\!-\!\epsilon_{{\bf k}})}{2\epsilon_{{\bf k}}+\Omega+i0^{+}}+\frac{(X_{{\bf k}}^{cd}\!-\!2\epsilon_{{\bf k}}^{\phantom{x}}\Omega\delta^{cd})\,\theta(\mu\!+\!\epsilon_{{\bf k}})}{2\epsilon_{{\bf k}}-\Omega+i0^{+}}
-\frac{(X_{{\bf k}}^{cd}\!-\!2\epsilon_{{\bf k}}^{\phantom{x}}\Omega\delta^{cd})\,\theta(\mu\!-\!\epsilon_{{\bf k}})}{2\epsilon_{{\bf k}}-\Omega+i0^{+}}-\frac{(X_{{\bf k}}^{cd}\!+\!2\epsilon_{{\bf k}}^{\phantom{x}}\Omega\delta^{cd})\,\theta(\mu\!+\!\epsilon_{{\bf k}})}{2\epsilon_{{\bf k}}+\Omega+i0^{+}} \Biggr\rbrace \nonumber
\end{eqnarray}
\end{widetext}
where:
\begin{eqnarray}
X_{{\bf k}}^{cd} &=& 2\epsilon_{{\bf k}}^{2}\delta^{cd} \\
&& + 2\left(\frac{v}{a_{0}}\right)^{2}\sin(a_{0}^{\phantom{x}}k^{a})\sin(a_{0}^{\phantom{x}}k^{b})(2\delta^{ac}\delta^{bd}-\delta^{ab}\delta^{cd}) \ . \nonumber
\end{eqnarray}
After some algebra, we can extract the imaginary part of the Raman susceptibility:
\begin{eqnarray}
&& -\frac{1}{\pi}\chi''({\bf q}\to0,\Omega) \approx \left(mva_{0}\right)^{2}\theta\left(\frac{|\Omega|}{2}-|\mu|\right) \\
&& \qquad\times \int\limits_{\textrm{1BZ}}\!\!\frac{d^{3}k}{(2\pi)^{3}}\,\delta(2\epsilon_{{\bf k}}-|\Omega|) \Biggl\lbrack -\sum_{a}(\hat{e}_{i}^{a}\hat{e}_{s}^{a})^{2}\,\sin^{2}(a_{0}k^{a}) \nonumber \\
&& \qquad\qquad\qquad\qquad~ + \left(\frac{2v}{a_{0}\Omega}\right)^{2}\left(\sum_{a}\hat{e}_{i}^{a}\hat{e}_{s}^{a}\sin^{2}(a_{0}k^{a})\right)^{\!\!2}\, \Biggr\rbrack \nonumber \ .
\end{eqnarray}
Further analytical progress is possible in the $|\Omega|\ll v/a_{0}$ limit, where we can also approximate $\sin(a_{0}k^{a})\approx a_{0}k^{a}$ and $\epsilon_{{\bf k}}\approx v|{\bf k}|$:
\begin{eqnarray}
&& -\frac{1}{\pi}\chi''\left({\bf q}\to0,|\Omega|\ll\frac{v}{a_{0}}\right) \approx \left(mva_{0}^{2}\right)^{2}\frac{4\Omega^{4}}{15(2\pi)^{2}(2v)^{5}} \nonumber \\
&& \qquad \times\theta\left(\frac{|\Omega|}{2}-|\mu|\right)\Biggl\lbrace-\sum_{a}(\hat{e}_{i}^{a}\hat{e}_{s}^{a})^{2}+\frac{1}{2}\sum_{a\neq b}(\hat{e}_{i}^{a}\hat{e}_{s}^{a})(\hat{e}_{i}^{b}\hat{e}_{s}^{b})\Biggr\rbrace \nonumber
\end{eqnarray}
Therefore, the frequency dependence of the Raman cross-section at small energy transfers is:
\begin{equation}
\frac{\partial^{2}\sigma}{\partial\Omega\partial\omega_{s}} \propto R(\Omega) \propto \Omega^{4}\:\theta\left(\frac{|\Omega|}{2}-|\mu|\right) \ ,
\end{equation}
and the polarization dependence in a simple cubic crystal obtains from:
\begin{eqnarray}
R &\propto& -(\hat{e}_{i}^{x}\hat{e}_{s}^{x})^{2}-(\hat{e}_{i}^{y}\hat{e}_{s}^{y})^{2}-(\hat{e}_{i}^{z}\hat{e}_{s}^{z})^{2} \\
&& +(\hat{e}_{i}^{x}\hat{e}_{s}^{x})(\hat{e}_{i}^{y}\hat{e}_{s}^{y})+(\hat{e}_{i}^{y}\hat{e}_{s}^{y})(\hat{e}_{i}^{z}\hat{e}_{s}^{z})+(\hat{e}_{i}^{z}\hat{e}_{s}^{z})(\hat{e}_{i}^{x}\hat{e}_{s}^{x}) \ . \nonumber \\[-0.05in] \nonumber
\end{eqnarray}
Both features derive from the high-energy part of the Weyl electron spectrum and do not reveal the chiral and relativistic nature of the Weyl nodes.

The high-energy spectrum captured here is tied to the considered microscopic model. Specifically, the vertex function $\gamma_{{\bf k}}$ picks the $\mathcal{O}(k^{3})$ term (order $l=3$) from the Taylor expansion of $\sin(k)$ in $H_{{\bf k}}$, which does not have any relation to the linear-$k$ Weyl node part. One could have just as well chosen a different dispersion with the lowest-order non-linear term in the $H_{{\bf k}}$ Taylor expansion being $l=2$ or $l=4$. This order affects the power of $k$ which is integrated out, and hence the power of frequency in $\chi(\Omega)$ when $\delta(2\epsilon_{{\bf k}}-|\Omega|)$ is applied. Consequently, the final power of $\Omega$ in the Raman scattering rate $R(\Omega)$ obtained with the effective mass approximation depends on the high-energy features of the electronic spectrum and has little or nothing to do with the presence of Weyl nodes. Based on this argument, one naively expects $R(\Omega)\sim\Omega^{2(l-2)+(d-1)}$ in $d$ dimensions. The result obtained with the effective mass approximation for graphene \cite{Wang2008b} is $R(\Omega)\sim\Omega$ and corresponds to $l=2$, $d=2$ which characterizes the energy dispersion modeled in that study.

The only somewhat reliable indication of Weyl nodes in the effective mass approximation appears to be a low frequency threshold $|\Omega|>\Omega_{0}=|2\mu|$  for the onset of Raman scattering. $\Omega_{0}$ is given by the smallest energy difference between the electron states at the same wavevector which are compatible via the selection rules. This energy difference can presumably be large in ordinary metals and band insulators, leading to $\Omega_{0}$ sizeable as a bandwidth or a gap between bands. A small $\Omega_{0}$ indicates the possibility of narrow-gap band-insulator or a nodal semimetal. If it is known by other means that the electronic spectrum has no gap in the relevant energy range, yet yields Raman scattering with a small $\Omega_{0}$, one can suspect a nodal semimetal. Otherwise, the frequency dependence of the Raman cross-section is not a clear indicator of the nodes.

% \bibliography{/home/dasko/Science/Bibliography/references}


\end{document}